\documentclass[a4paper,11pt]{article}
\pdfoutput=1 
\usepackage{jheppub} 
\usepackage[T1]{fontenc} 
\RequirePackage[displaymath]{lineno} 
\usepackage{epsfig}
\usepackage{graphicx}
\usepackage{dcolumn}
\usepackage{bm}
\usepackage{ltablex,booktabs}
\usepackage{overpic}
\usepackage{subfigure}
\usepackage{float}
\usepackage{color}
\usepackage{amsmath}
\usepackage{mathcomp}
\usepackage{mathrsfs}
\usepackage{multirow}
\usepackage{rotating}
\usepackage{amssymb}
\usepackage{gensymb}
\usepackage{amsmath}
\usepackage{tabularx}
\usepackage{epstopdf}
\usepackage{relsize}

\title{Amplitude analysis and branching fraction measurement of the decay \boldmath $D^{+} \to K_S^0\pi^+\pi^0\pi^0$}
\collaborationImg{\includegraphics[width=0.15\textwidth, angle=90]{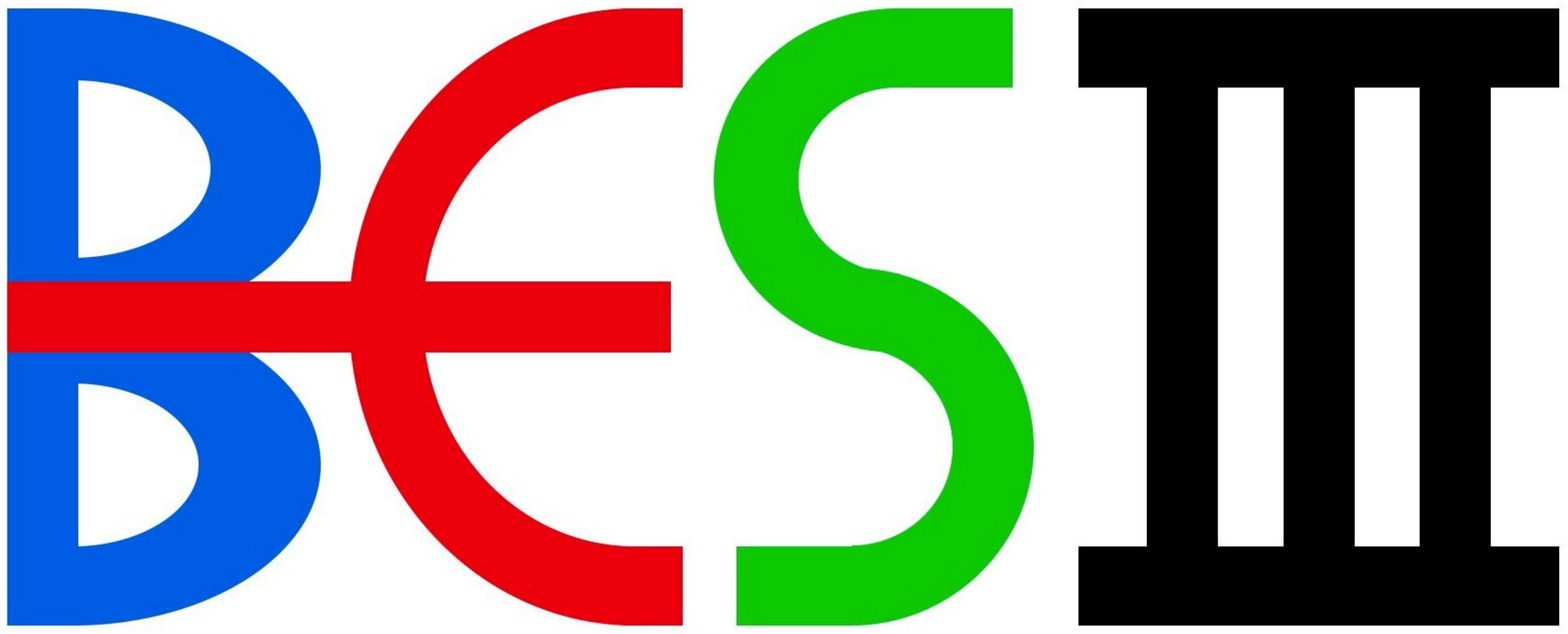}}
\collaboration{The BESIII Collaboration}

\date{\today}
\abstract{Using 2.93 $\rm{fb}^{-1}$ of $e^+e^-$ collision data collected with the BESIII detector at the center-of-mass energy 3.773\,GeV, we perform the first amplitude analysis of the decay $D^+\to K_S^0\pi^+\pi^0\pi^0$ and determine the relative magnitudes and phases of different intermediate processes. The absolute branching fraction of $D^+\to K_S^0\pi^+\pi^0\pi^0$ is measured to be $(2.888\pm0.058_{\rm stat.}\pm0.069_{\rm syst.})\%$. The dominant intermediate processes are $D^+\to K_S^0a_1(1260)^+(\to \rho^+\pi^0)$ and $D^+\to \bar{K}^{*0}\rho^+$, with branching fractions of $(8.66\pm1.04_{\rm stat.}\pm1.39_{\rm syst.})\!\times \!10^{-3}$ and $(9.70\pm0.81_{\rm stat.}\pm0.53_{\rm syst.})\!\times \!10^{-3}$, respectively.
}

\arxivnumber{}
\begin{document}
\maketitle
\flushbottom

\section{Introduction}
Hadrons containing a charm quark play an essential role in studies of the strong and weak interactions. The lightest charmed mesons, $D^{0(+)}$, can decay only through the weak interaction and their masses place them in the region where perturbative Quantum Chromodynamics is not applicable~\cite{Cheng:2010ry}. These facts do not significantly affect the theoretical prediction of leptonic and semileptonic decays but impose difficulties in hadronic decays~\cite{Ryd:2009uf}. Measurements of amplitudes and branching fractions~(BFs) of charmed meson hadronic decays could provide useful information about the underlying decay mechanism and help to improve theoretical calculations.

The Cabibbo-favored decay $D^{+} \to K_S^0\pi^+\pi^0\pi^0$ has been previously observed by BESIII with a BF of $(2.904\pm0.062_{\rm stat.}\pm0.087_{\rm syst.})\%$~\cite{BESIII:2022mji}. However, the corresponding detector efficiency was obtained by mixed-signal Monte Carlo (MC) samples, which can now be improved by using an amplitude model. An amplitude analysis of this four-body decay, compared to well-measured three-body decays~\cite{BESIII:2021dmo,BESIII:2014oag,FOCUS:2007mcb}, can also help us better understand the more complicated dynamics and substructures in the processes $D^+\to VV$ and $D^+\to AP$, where $V$, $A$, and $P$ denote vector, axial-vector and pseudoscalar mesons, respectively. The BF of $D^+\to \bar{K}^{*0}\rho^+$, which is a Cabibbo-favored $D^+\to VV$ process, can be measured more precisely in comparison to the previous MARK III result~\cite{MARK-III:1991fvi}. An amplitude analysis can provide inputs for polarization studies to check the reliability of different theoretical models~\cite{VV1}. Furthermore, measurements of $D^+\to AP$ decays are beneficial for our understanding of the nature of axial-vector mesons and offer global parameters in calculating the corresponding BFs~\cite{Guo:2018orw}. The difference between the production rates of $K_1(1270)$ and $K_1(1400)$ can be extracted, which provide key inputs to determine the mixing between these two mesons~\cite{Cheng:2011pb}.

With 2.93 $\rm{fb}^{-1}$ of $e^+e^-$ collision data collected by the BESIII detector at the center-of-mass energy $\sqrt{s} = 3.773$\,GeV, we present the first amplitude analysis of the decay $D^{+} \to K_S^0\pi^+\pi^0\pi^0$ and update the BF based on the corresponding amplitude model. The daughter particle $K_S^0$ is reconstructed by $\pi^+\pi^-$. Charge-conjugate states are implied throughout this paper. 

\section{Detector and data sets}
\label{sec:detector_dataset}
The BESIII detector is a magnetic spectrometer~\cite{ABLIKIM2010345,Ablikim_2020} located at the Beijing Electron Positron Collider (BEPCII)~\cite{Yu:IPAC2016-TUYA01}, which records symmetric $e^+e^-$ collisions in the center-of-mass energy range from 2.0 to 4.95~GeV, with a peak luminosity of $1 \times 10^{33}\;\text{cm}^{-2}\text{s}^{-1}$ achieved at $\sqrt{s} = 3.77\;\text{GeV}$.  The cylindrical core of the BESIII detector covers 93\% of the full solid angle and consists of a helium-based multilayer drift chamber (MDC), a plastic scintillator time-of-flight system (TOF), and a CsI(Tl) electromagnetic calorimeter (EMC), which are all enclosed in a superconducting solenoidal magnet providing a 1.0~T magnetic field. The solenoid is supported by an octagonal flux-return yoke with resistive plate counter muon identifier modules interleaved with steel. The charged-particle momentum resolution at 1.0~GeV/$c$ is 0.5\%, and the specific ionization energy loss (${\rm d}E/{\rm d}x$) resolution is 6\% for the electrons from Bhabha scattering. The EMC measures photon energies with a resolution of 2.5\%~(5\%) at 1 GeV in the barrel (end-cap) region. The time resolution in the TOF barrel region is 68~ps, while that in the end-cap region is 110~ps. 

Data samples corresponding to a total integrated luminosity of 2.93 f$\rm b^{-1}$ at $\sqrt{s}=3.773$~GeV are used in this analysis. This energy is slightly higher than the resonance peak of the $\psi(3770)$, which predominantly decays to $D^+D^-$ or $D^0\bar{D}^0$ pairs without any additional hadrons, thereby providing an ideal environment for studying $D$ meson decays with the double-tag~(DT) technique~\cite{MARK-III:1985hbd}. In this method, a single-tag (ST) candidate requires only one $D^{-}$ to be reconstructed via hadronic decays. In a DT candidate, both the $D^+$ and $D^-$ mesons are reconstructed, with the $D^+$ meson decaying to the signal mode $D^+\to K_S^0\pi^+\pi^0\pi^0$ and the $D^{-}$ meson decaying to one of the ST modes. 

Simulated inclusive MC samples are produced with a {\sc geant4}-based~\cite{GEANT4:2002zbu} MC simulation package, which includes the geometric description of the BESIII detector~\cite{Huang:2022wuo} and the detector response, and are used to determine detection efficiencies and to estimate backgrounds. The simulation models the beam energy spread and initial state radiation~(ISR) in the $e^+e^-$ annihilations with the generator {\sc kkmc}~\cite{Jadach:2000ir, Jadach:1999vf}. The inclusive MC samples consist of the production of $D\bar{D}$ pairs, the non-$D\bar{D}$ decays of the $\psi(3770)$, the ISR production of the $J/\psi$ and $\psi(3686)$ states, and the continuum processes incorporated in {\sc kkmc}. All particle decays are modelled with {\sc evtgen}~\cite{Lange:2001uf, EVTGEN2} using BFs either taken from the Particle Data Group (PDG)~\cite{PDG}, when available, or otherwise estimated with {\sc lundcharm}~\cite{Chen:2000tv, LUNDCHARM2}. Final state radiation from charged final state particles is incorporated using {\sc photos}~\cite{PHOTOS}. 

\section{Event selection}
\label{ST-selection}
The $D^\pm$ candidates are constructed from individual $\pi^\pm$, $\pi^0$, $K^\pm$, and $K_S^0$ mesons with the following selection criteria, which are the common requirements for both the amplitude analysis and {{BF}} measurement. Further requirements are discussed in Sec.~\ref{AASelection} and Sec.~\ref{BFSelection}, respectively.

All charged tracks detected in the MDC must satisfy $|$cos$\theta|<0.93$, where $\theta$ is defined as the polar angle with respect to the $z$-axis, which is the symmetry axis of the MDC. For charged tracks not originating from $K_S^0$ decays, the distance of closest approach to the interaction point (IP) is required to be less than 10\,cm along the $z$-axis, $|V_{z}|$, and less than 1\,cm in the transverse plane, $|V_{xy}|$. Particle identification (PID) for charged tracks combines the measurements of the ${\rm d}E/{\rm d}x$ in the MDC and the flight time in the TOF to form probabilities $\mathcal{L}(h)~(h=K,\pi)$ for each hadron ($h$) hypothesis. The charged kaons and pions are identified by comparing the likelihoods for the kaon and pion hypotheses, $\mathcal{L}(K)>\mathcal{L}(\pi)$ and $\mathcal{L}(\pi)>\mathcal{L}(K)$, respectively. 

Each $K_{S}^0$ candidate is reconstructed from two oppositely charged tracks satisfying \mbox{$|V_{z}|<$ 20~cm}. The two charged tracks are assigned as $\pi^+\pi^-$ without imposing PID. They are constrained to originate from a common vertex and are required to have an invariant mass within $|M_{\pi^{+}\pi^{-}} - m_{K_{S}^{0}}|<$ 12~MeV$/c^{2}$, where $m_{K_{S}^{0}}$ is the known $K^0$ mass~\cite{PDG}. The decay length of the $K^0_S$ candidate is required to be greater than twice its resolution.
 
Photon candidates are selected using the EMC showers. The deposited energy of each shower in the barrel region~($|\!\cos \theta|< 0.80$) and in the end-cap region~($0.86 <|\!\cos \theta|< 0.92$) must be greater than 25 MeV and 50 MeV, respectively. To exclude showers that originate from charged tracks, the angle subtended by the EMC shower and the position of the closest charged track at the EMC must be greater than 10 degrees as measured from the interaction point. The difference between the EMC time and the event start time is required to be within \mbox{[0, 700]~ns} to suppress electronic noise and showers unrelated to the event.

The $\pi^0$ candidates are reconstructed from photon pairs with invariant masses in the range $[0.115, 0.150]$~GeV/$c^{2}$, which corresponds to about three times the standard deviation of the invariant mass resolution. We require that at least one photon comes from the barrel region of the EMC to improve the resolution. Furthermore, the $\pi^0$ candidates are constrained to the known $\pi^0$ mass~\cite{PDG} via a kinematic fit to improve their energy and momentum resolution.

Two variables, the beam-constrained mass $M_{\rm BC}$ and the energy difference $\Delta E$, are used to identify the $D^\pm$ mesons:
\begin{eqnarray}
\begin{aligned}
    M_{\rm BC}&=\sqrt{E^2_{\rm beam}/c^4-|\vec{p}_{D^\pm}|^2/c^2},\\
	\Delta{E} &= E_{D^\pm}-E_{\rm beam}, \label{eq:mbc}
\end{aligned}
\end{eqnarray}
where $E_{\rm beam}$ is the calibrated beam energy, and $\vec{p}_{D^\pm}$ and $E_{D^\pm}$ are the total reconstructed momentum and energy of the $D^\pm$ candidate, respectively. The $D^\pm$ signals will appear as a peak at the known $D^\pm$ mass~\cite{PDG} in the $M_{\rm BC}$ distribution and as a peak at zero in the $\Delta{E}$ distribution. If multiple DT candidates exist in one event, the candidate with the minimum quadratic sum of $\Delta{E}$ from the two $D^\pm$ mesons ($\Delta{E}_{\rm tag}^2+\Delta{E}_{\rm sig}^2$) is retained.

\section{Amplitude analysis}
\label{Amplitude-Analysis}
\subsection{Further selection criteria}
\label{AASelection}
To increase the signal purity for amplitude analysis, only one ST mode \mbox{$D^-\to K^+ \pi^- \pi^-$} is used due to its high statistics and low background level. We require that $\Delta{E}\!\in\!\left[-0.025, 0.020\right]$ ($\left[-0.040, 0.020\right]$)~GeV must be satisfied for the tag~(signal) side. The following dedicated selection criteria are imposed on the signal candidates and will not be used in the BF measurement. 

\begin{figure*}[!htbp]
  \centering
    \includegraphics[width=6.6cm]{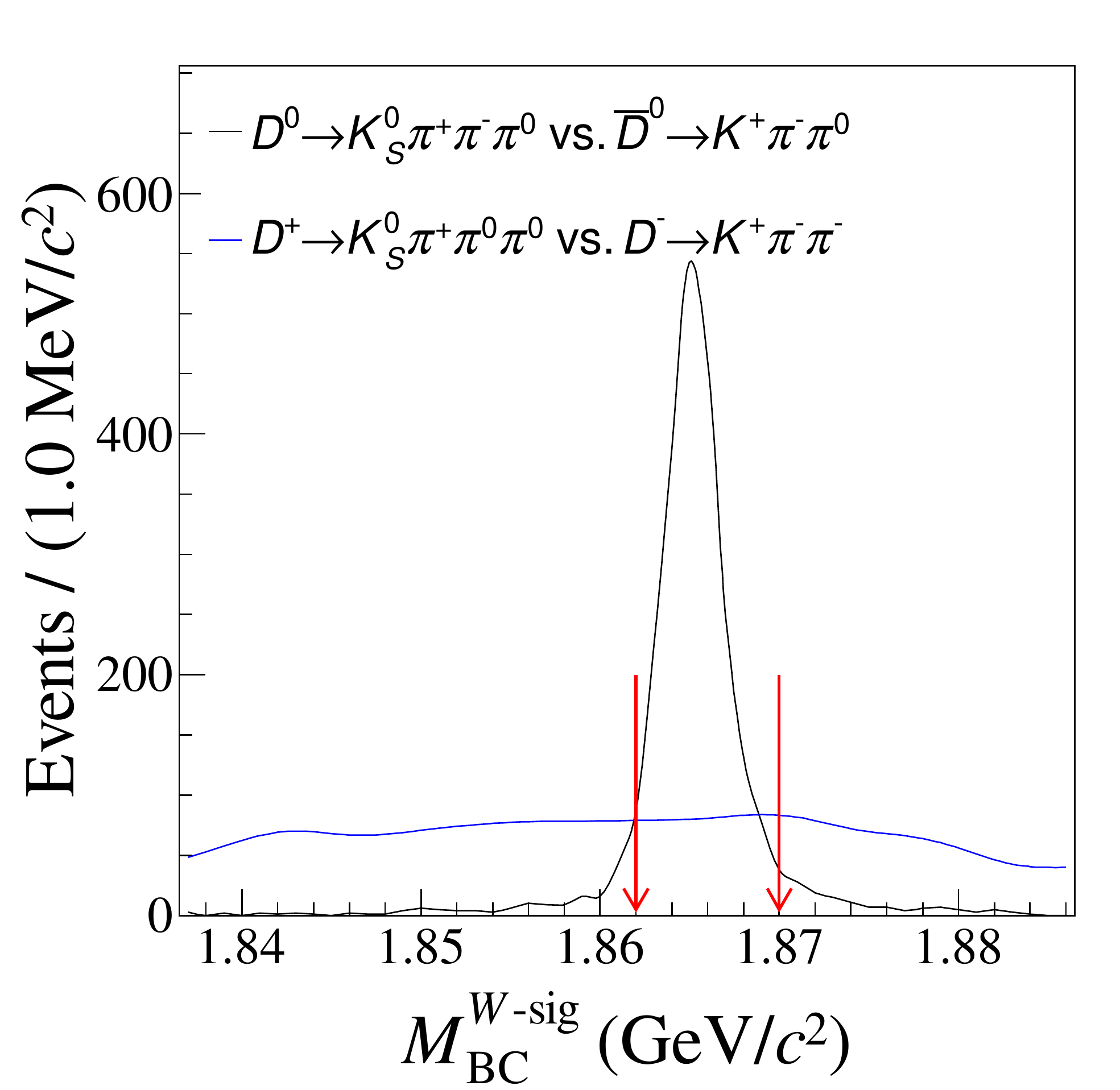}
    \includegraphics[width=6.6cm]{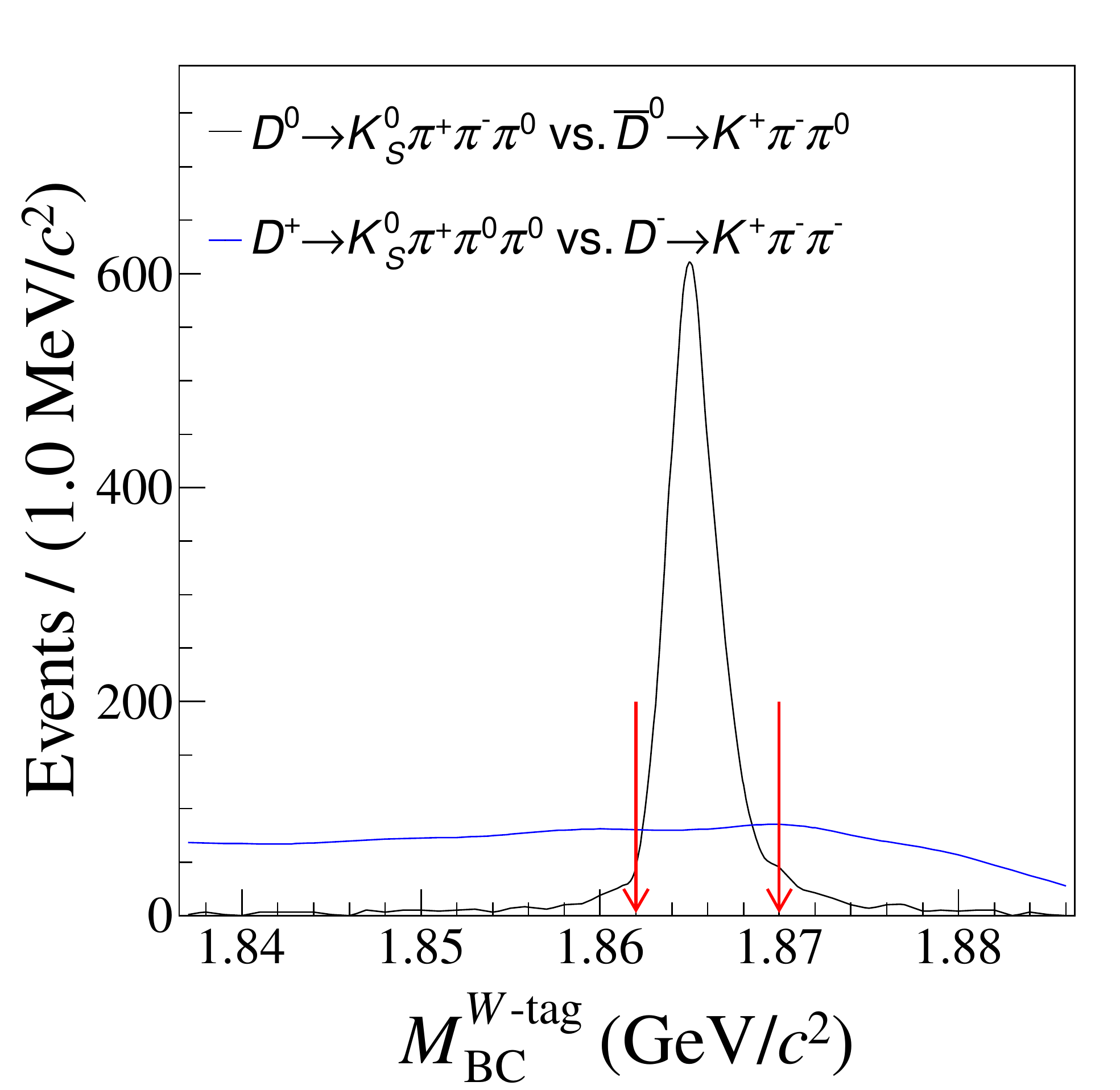}
  \caption{
    The $M_{\rm BC}^W$ distributions for miscombined background and the signal process from MC simulation. The left is for signal side, and the right is for tag side. The black and blue lines are miscombined background and the signal process, respectively. The pairs of red arrows indicate the related veto region. 
  } \label{fig:misp}
\end{figure*}

A $K_S^0$ mass veto, $M_{\pi^0\pi^0}\!\notin\!\left[0.46, 0.52\right]$~GeV/$c^2$, is used to suppress the dominant background  from $D^+\to K_S^0K_S^0\pi^+$ in which one of the two $K_S^0$ mesons decays to $\pi^0\pi^0$. Another source of background comes from the process $D^0\to K_S^0\pi^+\pi^-\pi^0$ versus $\bar{D}^0\to K^+\pi^-\pi^0$, which can be miscombined to fake the signal process $D^+\to K_S^0\pi^+\pi^0\pi^0$ versus $D^- \to K^+\pi^-\pi^-$ by exchanging the $\pi^-$ from $D^0$ decay and the $\pi^0$ from $\bar{D}^0$ decay. We reconstruct this background and calculate the wrong beam-constrained mass $M_{\rm BC}^W$ and the wrong energy difference $\Delta E^{W}$ according to the $D^0\bar{D}^0$ decay mode. For multiple miscombined candidates, we use the minimum quadratic sum of $\Delta E^W$ to select the ``best background'' event. Figure~\ref{fig:misp} shows the $M_{\rm BC}^W$ distribution for this background and the signal process from MC simulation. The background will form a peak at the known $D^0$ mass~\cite{PDG} while the distribution for signal is flat. Therefore, it is excluded by rejecting events which simultaneously satisfy $1.862<M_{\rm BC}^W<1.870$~GeV/$c^2$ for both the tag and the signal sides.

\begin{figure*}[!htbp]
  \centering
    \includegraphics[width=6.6cm]{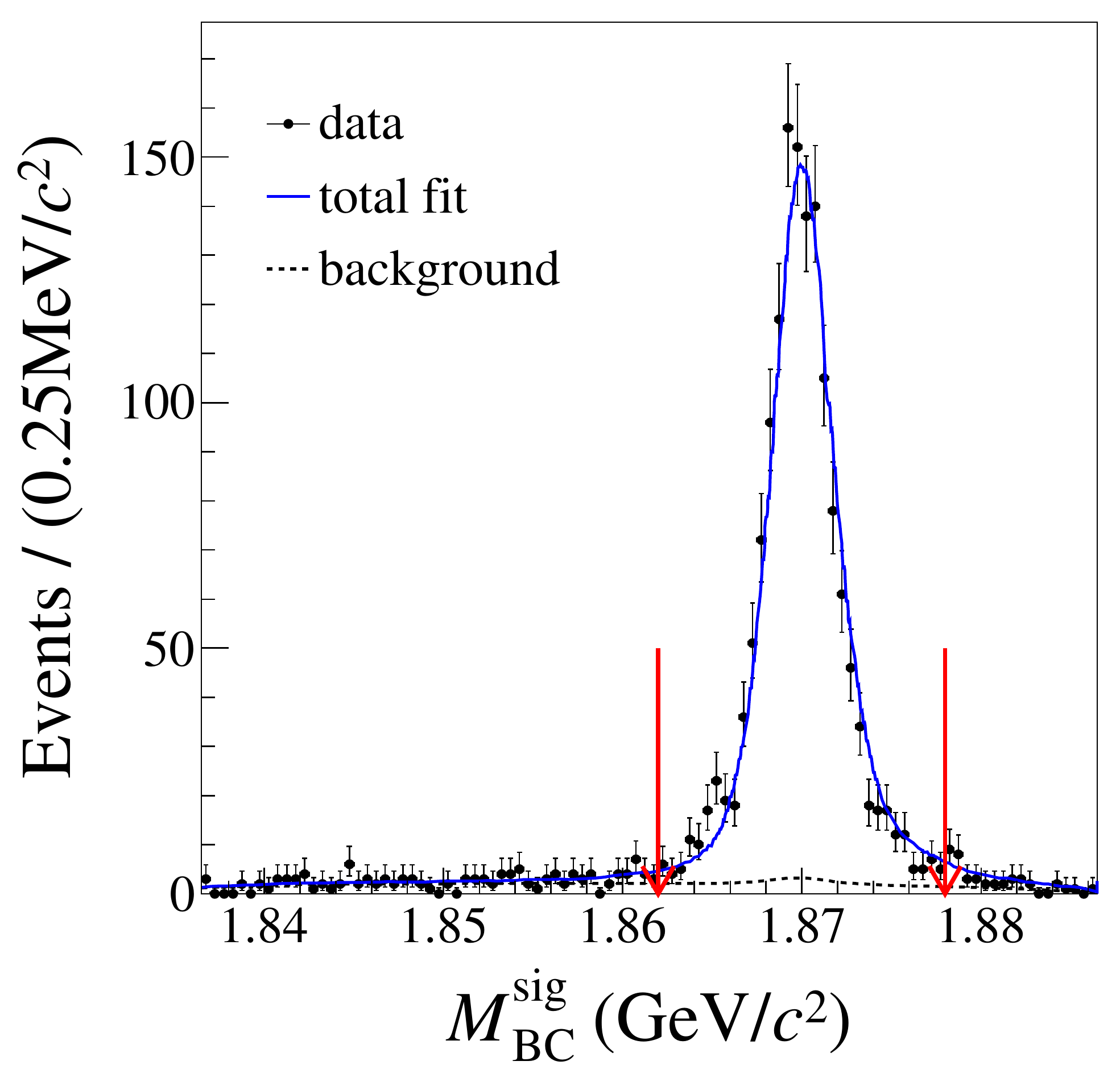}
    \includegraphics[width=6.6cm]{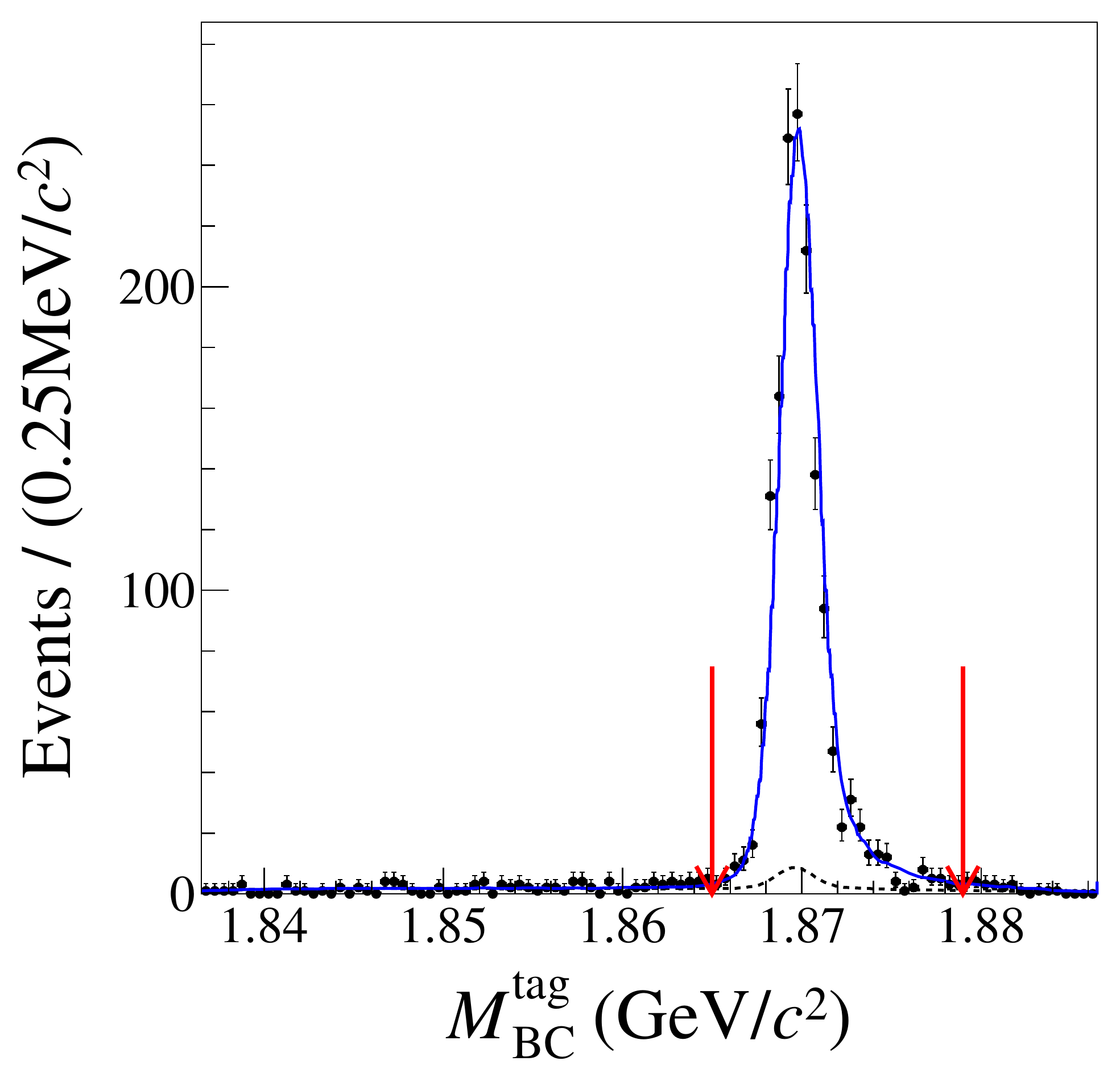}
  \caption{
    The projections of the 2D fit on $M_{\rm BC}^{\rm sig}$~(left) versus $M_{\rm BC}^{\rm tag}$~(right). The black points with error bars are data. The blue solid and black dashed lines are the total fits and the backgrounds, respectively. The pairs of red arrows indicate the signal region, $\left[1.862, 1.878\right]$~GeV/$c^2$ for the signal side and $\left[1.865, 1.879\right]{\rm GeV}/c^2$ for the tag side.
  } \label{fig:pwa_purity}
\end{figure*}

We perform a four-constraint kinematic fit to ensure that all events land within the phase space boundary. The invariant masses of the signal $D^+$ candidate, the $K_S^0$, and the two $\pi^0$s are constrained to the corresponding known masses~\cite{PDG}. The updated four-momenta of the final state particles from the kinematic fit are used to perform the amplitude analysis. After applying all selection criteria, we perform an unbinned two-dimensional (2D) maximum likelihood fit (see appendix~\ref{app:2dfit}) to the distribution $M_{\rm BC}^{\rm tag}$ versus $M_{\rm BC}^{\rm sig}$ to estimate the signal purity, as shown in figure~\ref{fig:pwa_purity}. There are 1,458 events remaining in the signal region for amplitude analysis with a purity of $(96.86 \pm 0.46)\%$.

\subsection{Fit method}
\label{AAfitmethod}
The amplitude analysis of $D^{+}\to K_S^0\pi^+\pi^0\pi^0$ is performed by an unbinned maximum likelihood fit. The likelihood function $\mathcal{L}$ is constructed by adding the background probability density function (PDF) to the signal PDF incoherently. After taking the logarithm, the combined PDF can be written as
\begin{equation}
  \ln{\mathcal{L}} = \begin{matrix}\sum\limits_{k}^{N_{\rm data}} \ln [\omega_{\rm sig}f_{S}(p_{j}^{k})+(1-\omega_{\rm sig})f_B(p^k_j)]\end{matrix},
  \label{eq:likefinal}
\end{equation}
where $k$ indicates the $k^{\rm th}$ event in the data sample, $N_{\rm data}$ is the number of surviving events, $p_j$ denotes the four-momenta of the $j^{th}$ final state particles, $f_S$($f_B$) is the signal (background) PDF and $\omega_{sig}$ is the signal purity discussed in Sec.~\ref{AASelection}.

The signal PDF is given by 
\begin{equation}
f_S(p_j) = \frac{\epsilon(p_j)|\mathcal M(p_j)|^2R_4(p_j)}{\int \epsilon(p_j)|\mathcal M(p_j)|^2R_4(p_j)dp_j},
\label{pwa:pdf}
\end{equation}
where $\epsilon(p)$ is the detection efficiency and $R_4(p)$ is the four-body phase space. The total amplitude $\mathcal{M}$ is treated within the isobar model, which uses the coherent sum of amplitudes of intermediate processes, given by $\mathcal M(p) = \sum{c_n\mathcal A_n(p)}$, where $c_n = \rho_ne^{i\phi_n}$ and $\mathcal A_n(p)$ are the complex coefficient and the amplitude for the $n^{\rm th}$ intermediate process, respectively. The magnitude $\rho_n$ and phase $\phi_n$ are free parameters in the fit. We use covariant tensors to construct amplitudes, which are written as
\begin{equation}
  \mathcal A_n(p_j) = P_n^1P_n^2S_nX_n^1X_n^2X_n^{D^+},
\end{equation}
where $S_n$ and { $X_n^{1,2}(X_n^{D^+})$} are the spin factor and the Blatt-Weisskopf barriers for the intermediate resonances (the $D^+$ meson), respectively. The propagators of the two resonances, which describe the corresponding lineshapes, are indicated by $P_n^{1,2}$ . Their specific forms will be introduced in Sec.~\ref{Blatt-Weisskopfbarriers}-\ref{Spinfactors}.

For the amplitude of $D^-$ decays, we define the $C\!P$ conjugate phase space $\bar{p_j}$ which is mapped to $p_j$ by the interchange of final state charges and the reversal of three-momenta, and assume $C\!P$ conservation in the $D^\pm$ decay. Then we get
\begin{equation}
\mathcal M_{D^-}(p_j)=\sum_n{c_n\bar{\mathcal A}_n(p_j)}=\sum_n{c_n\mathcal A_n(\bar{p}_j)}.
\end{equation}

The background PDF is given by
\begin{equation}
    f_B(p_j)=\frac{B(p_j)R_4(p_j)}{\int{\epsilon(p_j)B_{\epsilon}(p_j)R_4(p_j)}dp_j},	
  \label{bkglikelihood}
\end{equation}
where $B_{\epsilon}(p_j)=B(p_j)/\epsilon(p_j)$ is the efficiency-corrected background shape. The shape of the background in data is modeled by the background events in the signal region derived from the inclusive MC samples. The invariant mass distributions of events outside the signal region show good agreement between data and MC simulation, thus validating the description from the inclusive MC samples. We have also examined the distributions of the background events of the inclusive MC samples inside and outside the signal region. Generally, they are compatible with each other within statistical uncertainties. The background shape $B(p)$ is modeled using a kernel estimation method~\cite{CRANMER2001198} implemented in RooFit~\cite{RooNDKeysPDF} to model the distribution of an input dataset as a superposition of Gaussian kernels.

In the numerator of Eq.~(\ref{pwa:pdf}), the $\epsilon(p_j)$ and $R_4(p_j)$ terms are independent of the fitted variables, so they are regarded as constants in the fit. As a consequence, the log-likelihood becomes
\begin{equation}
  \ln\mathcal{L}=\sum\limits_k^{N_{\rm data}}\ln\bigg[\omega_{\rm sig}\frac{|\mathcal M(p_j)|^2}{\int \epsilon(p_j)|\mathcal M(p_j)|^2R_4(p_j)dp_j}+(1-\omega_{\rm sig})\frac{B_\epsilon(p_j)}{\int{\epsilon(p_j)B_\epsilon(p_j)R_4(p_j)}dp_j}\bigg].
  \label{likelihoodfinal}
\end{equation}

The normalization integrals of signal and background are evaluated by MC integration,
\begin{equation}
\begin{aligned}
  \int \epsilon(p_j)|\mathcal M(p_j)|^2R_4(p_j)dp_j &\propto \frac{1}{N_{\rm MC}}\sum^{N_{\rm MC}}_{k_{\rm MC}}\frac{|\mathcal M(p^{k_{\rm MC}}_j)|^2}{|\mathcal M^{\rm gen}(p^{k_{\rm MC}}_j)|^2}, \\
  \int \epsilon(p_j)B_\epsilon(p_j)R_4(p_j)dp_j &\propto \frac{1}{N_{\rm MC}}\sum^{N_{\rm MC}}_{k_{\rm MC}}\frac{B_\epsilon(p_j^{k_{\rm MC}})}{|\mathcal M^{\rm gen}(p^{k_{\rm MC}}_j)|^2},
\end{aligned}
\end{equation}
where $k_{\rm MC}$ is the index of the $k^{\rm th}$ event of the MC sample and $N_{\rm MC}$ is the number of selected MC events. The $\mathcal M^{\rm gen}(p_j)$ is the signal PDF used to generate the MC samples in MC integration.

Tracking, PID, as well as $\pi^0$ and $K_S^0$ reconstruction efficiency differences between data and MC simulation are corrected by multiplying the weight of the MC events by a factor $\gamma_{\epsilon}$, which is calculated as
\begin{equation}
  \gamma_{\epsilon}(p_j) = \prod_{n} \frac{\epsilon_{n,\rm data}(p_j)}{\epsilon_{n,\rm MC}(p_j)},
  \label{pwa:gamma}
\end{equation}
where $n$ refers to tracking, PID, $\pi^0$ reconstruction or $K_S^0$ reconstruction, $\epsilon_{n,\rm data}(p_j)$ and $\epsilon_{n,\rm MC}(p_j)$ are their  efficiencies as a function of the momenta of the daughter particles for data and MC simulation, respectively. The specific values of these efficiencies are obtained using different control samples, more detailed information will be given as part of the systematic uncertainty studies for the BF measurement. By weighting each signal MC event with $\gamma_{\epsilon}$, the MC integration is modified to be
\begin{equation}
  \int \epsilon(p_j)|\mathcal M(p_j)|^2R_4(p_j)dp_j \propto\frac{1}{N_{\rm MC}}\sum^{N_{\rm MC}}_{k_{\rm MC}}\frac{\gamma_{\epsilon}(p_j^{k_{\rm MC}})|\mathcal{M}(p_j^{k_{\rm MC}})|^2}{|\mathcal{M}^{\rm gen}(p_j^{k_{\rm MC}})|^2}.\label{likelihood3}
\end{equation}
\begin{equation}
  \int \epsilon(p_j)B_\epsilon(p_j)R_4(p_j)dp_j \propto\frac{1}{N_{\rm MC}}\sum^{N_{\rm MC}}_{k_{\rm MC}}\frac{\gamma_{\epsilon}(p_j^{k_{\rm MC}})B_\epsilon(p_j^{k_{\rm MC}})}{|\mathcal{M}^{\rm gen}(p_j^{k_{\rm MC}})|^2}.\label{likelihood4}
\end{equation}

\subsubsection{Blatt-Weisskopf barriers}
\label{Blatt-Weisskopfbarriers}
For a decay process $a \to b\ c$, the Blatt-Weisskopf barrier {{factors}}~\cite{PhysRevD.104.012016} depend on the angular momentum $L$ and the momentum $q$ of the final-state particle $b$ or $c$ in the rest system of $a$. They are taken as
\begin{equation}
\begin{aligned}
  &X_{L=0}(q)=1,\\
  &X_{L=1}(q)=\sqrt{\frac{z_0^2+1}{z^2+1}},\\
  &X_{L=2}(q)=\sqrt{\frac{z_0^4+3z_0^2+9}{z^4+3z^2+9}}, \label{xl}
\end{aligned}
\end{equation}
where $z = qR$ and $z_0 = q_0R$. The effective radius of barrier $R$ is fixed to be 3.0 $(\rm {GeV}/c)^{-1}$ for the intermediate resonances and 5.0 $(\rm {GeV}/c)^{-1}$ for the $D^+$ meson. The momentum $q$ is given by
\begin{equation}
q = \sqrt{\frac{(s_a+s_b-s_c)^2}{4s_a}-s_b}, \label{q2}
\end{equation}
where $s_a, s_b,$ and $s_c$ are the invariant mass squared of particles $a,\ b$ and $c$, respectively. The value of $q_0$ is that of $q$ when $s_a = m^2_a$, where $m_a$ is the mass of particle $a$.

\subsubsection{Propagator}
\label{Propagator}
The intermediate resonances $a_1(1260)^+$, $\bar{K}^{*0}$, $K_1(1270)^+$ and $\bar{K}_1(1400)^0$ are parameterized with the relativistic Breit-Wigner (RBW) formulas,
\begin{equation}
\begin{aligned}
		&P(m) = \frac{1}{(m^2_0-m^2)-im_0\Gamma(m)}, \\ 
		&\Gamma(m)=\Gamma_0\left(\frac{q}{q_0}\right)^{2L+1}\Big(\frac{m_0}{m}\Big)\left(\frac{X_L(q)}{X_L(q_0)}\right)^2,  \label{propagator}
\end{aligned}
\end{equation}
where $m^2$ is the invariant mass squared of the daughter particles of the intermediate resonances, and $m_0$ and $\Gamma_0$ are the mass and width of the intermediate resonance, which are fixed to their known values~\cite{PDG}. The energy-dependent width is denoted by $\Gamma(m)$.  

The $a_1(1260)^+$ decays through a quasi-three-body process $r\to a\,b\,c$, whose energy-dependent width is more complicated and has no analytic expression in general. Therefore, we integrate the transition amplitude squared over the three-body phase space $R_3$~\cite{dArgent:2017gzv}
\begin{equation}
\Gamma_{r\to a\,b\,c}(m)=\frac{1}{2\sqrt{s}}\int\lvert \mathcal A_{r\to a\,b\,c}\rvert^{2}dR_{3}.
\label{width}
\end{equation}

\begin{figure*}[!htbp]
  \centering
    \includegraphics[width=7.2cm]{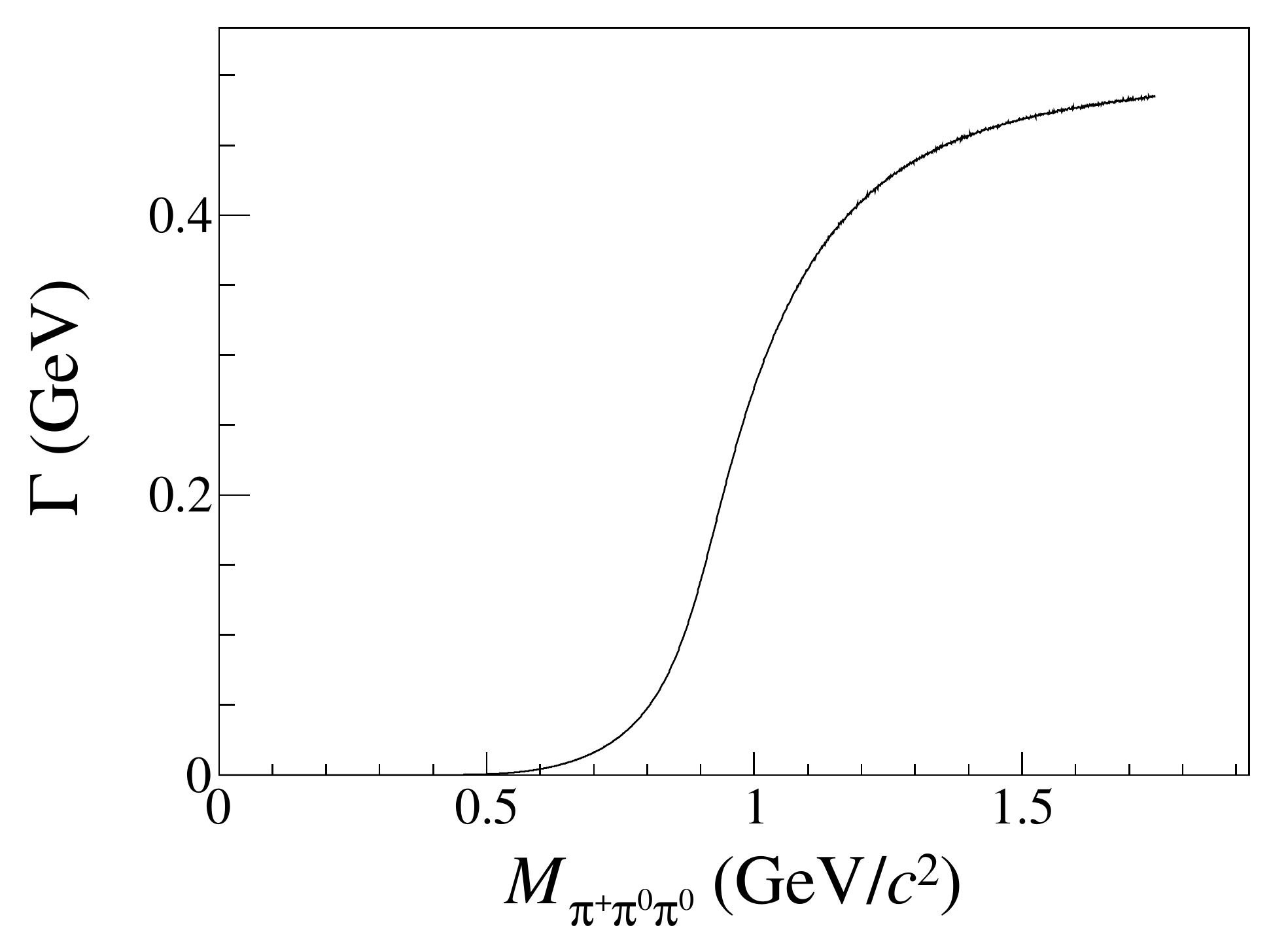}
  \caption{The energy-dependent width for the $a_1(1260)^+$.} 
  \label{fig:width}
\end{figure*}

The three-body amplitude $\mathcal A_{r\to a\,b\,c}$ can be parameterized similarly to the four-body amplitude and is obtained from the amplitude analysis of this work. Figure~\ref{fig:width} shows the energy-dependent width for the $a_1(1260)^+$ resonance. 

The $\rho^+$ meson is parameterized using the Gounaris-Sakurai (GS) line shape~\cite{Gounaris:1968mw}, which is given by
\begin{equation}
P_{\rm GS}(m)=\frac{1+d\frac{\Gamma_0}{m_0}}{(m^2_0-m^2)+f(m)-im_0\Gamma(m)},
\end{equation}
where
\begin{equation}
	f(m)=\Gamma_0\frac{m^2_0}{q^3_0}\left[q^2(h(m)-h(m_0))+(m^2_0-m^2)q^2_0\frac{dh}{d(m^2)}\Big|_{m^2=m^2_0}\right]
\end{equation}
and the function $h(m)$ is defined as
\begin{equation}
h(m)=\frac{2}{\pi}\frac{q}{m}{\rm ln}\left(\frac{m+2q}{2m_{\pi}}\right),
\end{equation}
with
\begin{equation}
\frac{dh}{d(m^2)}\Big|_{m^2=m_0^2} = h(m_0)[(8q^2_0)^{-1}-(2m^2_0)^{-1}]+(2\pi m^2_0)^{-1},
\end{equation}
where $m_{\pi}$ is the known mass of the $\pi$~\cite{PDG}, and the normalization condition at $P_{\rm GS}(0)$ fixes the parameter $d=\frac{f(0)}{\Gamma_0m_0}$. It is found to be
\begin{equation}
d=\frac{3}{\pi}\frac{m^2_{\pi}}{q^2_0}{\rm ln}\left(\frac{m_0+2q_0}{2m_{\pi}}\right)+\frac{m_0}{2\pi q_0}-\frac{m^2_{\pi}m_0}{\pi q^3_0}.
\end{equation}

The $f_0(980)$ is parameterized with the Flatt\'e formula~\cite{BES:2004twe}:
\begin{equation}
	P_{f_0(980)}=\frac{1}{M_{f_0(980)}^2-m^2-i(g_{\pi\pi}\rho_{\pi\pi}(m^2)+g_{K\bar{K}}\rho_{K\bar{K}}(m^2))},
\end{equation}
where $g_{\pi\pi,K\bar{K}}$ are the coupling constants to individual final states. The parameters are fixed to be $g_{\pi\pi}=(0.165{ \pm 0.010 \pm 0.015)} {\rm GeV^2}/c^4$, $g_{K\bar{K}}=(4.21{ \pm 0.25 \pm 0.21)}g_{\pi\pi}$ and $M = 0.965$~GeV/$c^2$, as reported in Ref.~\cite{BES:2004twe}. The Lorentz invariant phase space factors $\rho_{\pi\pi}(s)$ and $\rho_{K\bar{K}}(s)$ are given by
\begin{equation}
\begin{aligned}
	\rho_{\pi\pi}&=\frac{2}{3}\sqrt{1-\frac{4m^2_{\pi^\pm}}{m^2}}+\frac{1}{3}\sqrt{1-\frac{4m^2_{\pi^0}}{m^2}},\\
	\rho_{K\bar{K}}&=\frac{1}{2}\sqrt{1-\frac{4m^2_{K^\pm}}{m^2}}+\frac{1}{2}\sqrt{1-\frac{4m^2_{K^0}}{m^2}}.
\end{aligned}
\end{equation}

The resonance $f_0(500)$ is parameterized with the formula given in Ref.~\cite{BUGG199659}:
\begin{equation}
\begin{aligned}
	P_{f_0(500)}=\frac{1}{m_0^2-m^2-im_0\Gamma_{\rm tot}(m)},
\end{aligned}
\end{equation}
where $\Gamma_{\rm tot}(m)$ is decomposed into two parts:
\begin{equation}
\begin{aligned}
	\Gamma_{\rm tot}(m)=g_{\pi\pi}\frac{\rho_{\pi\pi}(m)}{\rho_{\pi\pi}(m_0)}+g_{4\pi}\frac{\rho_{4\pi}(m)}{\rho_{4\pi}(m_0)}
\end{aligned}
\end{equation}
and
\begin{equation}
\begin{aligned}
	g_{\pi\pi}=(b_1+b_2m^2)
   \left( \frac{m^2-m_\pi^2/2}{m_0^2-m_\pi^2/2} \right) \; e^{(m_0^2-m^2)/a},
\end{aligned}
\end{equation}
where $\rho_{\pi\pi}$ and $\rho_{4\pi}$ are the phase space of the $\pi^+\pi^-$ and $4\pi$ systems, respectively. They are approximated by
\begin{equation}
\begin{aligned}
	\rho_{\pi\pi}=\sqrt{\left(1-\frac{4m_\pi^2}{m^2}\right)},\ \rho_{4\pi}  = \sqrt{\left(1-\frac{16m_\pi^2}{m^2}\right)}/(1+e^{3.5(2.8-m^2)}),
\end{aligned}
\end{equation}
with the parameters fixed to the values given in Ref.~\cite{Pelaez:2015qba}.

The $K\pi$ S-wave modeled by the LASS parameterization~\cite{Aston:1987ir} is described by a $K^*_0(1430)$ Breit-Wigner together with an effective range non-resonant component with a phase shift. It is given by
\begin{equation}
A(m) = F{\rm sin}\delta_F e^{i\delta_F} + R{\rm sin}\delta_R e^{i\delta_R}e^{i2\delta_F},
\end{equation}
with
\begin{equation}
\begin{aligned}
&\delta_F = \phi_F + {\rm cot}^{-1}\Big[\frac{1}{aq}+\frac{rq}{2}\Big],\\
&\delta_R = \phi_R + {\rm tan}^{-1}\Big[\frac{M\Gamma(m_{K\pi})}{M^2-m^2_{K\pi}}\Big],
\end{aligned}
\end{equation}
where the parameters $F(\phi_F)$ and $R(\phi_R)$ are the magnitudes (phases) for non-resonant state and resonance terms, respectively. The parameters $a$ and $r$ are the scattering length and effective interaction length, respectively. We fix these parameters ($M, \Gamma, F, \phi_F, R, \phi_R, a, r$) to the results obtained from the amplitude analysis to a sample of $D^0 \to K_S^0\pi^+\pi^-$ by the \mbox{\slshape B\kern-0.1em{\smaller A}\kern-0.1em B\kern-0.1em{\smaller A\kern-0.2em R}} and Belle experiments~\cite{PhysRevD.98.112012}; these parameters are summarised in table~\ref{LASSpa}.

\begin{table}[htbp]
  \begin{center}
    \begin{tabular}{| l c |}
      \hline
      $M\ ({\rm GeV}/c^2)$ &1.441 $\pm$ 0.002\\
      $\Gamma\ ({\rm GeV})$ &0.193 $\pm$ 0.004\\
      $F$ &0.96 $\pm$ 0.07\\
      $\phi_F\ (^\circ)$ &0.1 $\pm$ 0.3\\
      $R$ &{1(fixed)}\\
      $\phi_R\ (^\circ)$ &-109.7 $\pm$ 2.6\\
      $a\ ({\rm GeV}/c)^{-1}$ &0.113 $\pm$ 0.006\\
      $r\ ({\rm GeV}/c)^{-1}$ &-33.8 $\pm$ 1.8\\
      \hline
    \end{tabular}
  \end{center}
  \caption{The $K\pi$ S-wave parameters, obtained from the amplitude analysis of $D^0 \to K_S^0\pi^+\pi^-$ by the \mbox{\slshape B\kern-0.1em{\smaller A}\kern-0.1em B\kern-0.1em{\smaller A\kern-0.2em R}} and Belle experiments~\cite{PhysRevD.98.112012}. Uncertainties are statistical only.}
  \label{LASSpa}
\end{table}

\subsubsection{Spin factors}
\label{Spinfactors}
Due to the limited size of phase space, we only consider states with angular momenta below three. For a two-body decay, $a \to bc$, we use the notation $p_a$, $p_b$ and $p_c$ as the momenta of particles $a$, $b$ and $c$, respectively, and let $r_a = p_b - p_c$ be the break-up four-momentum. The spin projection operators are defined as
\begin{equation}
\begin{aligned}
&P^0(a) = 1,&(S~{\rm wave})\\
&P^{(1)}_{\mu\nu}(a)=-g_{\mu\nu}+\frac{p_{a\mu}p_{a\nu}}{p^2_a},&(P~{\rm wave})\\
&P^{(2)}_{\mu_1\mu_2\nu_1\nu_2}(a)=\frac{1}{2}(P^{(1)}_{\mu_1\nu_1}P^{(1)}_{\mu_2\nu_2}+P^{(1)}_{\mu_1\nu_2}P^{(1)}_{\mu_2\nu_1})-\frac{1}{3}P^{(1)}_{\mu_1\mu_2}P^{(1)}_{\nu_1\nu_2}.&(D~{\rm wave})
\end{aligned}
\end{equation}
The covariant tensors are given by
\begin{equation}
\begin{aligned}
&\tilde{t}^{(0)}(a) = 1,&(S~{\rm wave})\\
&\tilde{t}^{(1)}_\mu(a) = -P^{(1)}_{\mu\nu}(a)r_a^{\nu},&(P~{\rm wave})\\
&\tilde{t}^{(2)}_{\mu\nu}(a) = P^{(2)}_{\mu\mu_1\nu\nu_1}(a)r_a^{\mu_1}r_a^{\nu_1}.&(D~{\rm wave})
\end{aligned}
\end{equation}
The spin factors $S(p)$ used in this work are constructed from the spin projection operators and pure orbital angular-momentum covariant tensors and are listed in table~\ref{table:spin_factors}.
\begin{table}[hbtp]
 \begin{center}
\begin{tabular}{| l c |}
\hline
 Decay chain& $S(p)$ \\
 \hline
$D^+[S]\to V_1V_2$ & $\tilde{t}^{(1)\mu}(V_1)\tilde{t}^{(1)}_\mu(V_2)$ \\
$D^+[P]\to V_1V_2$ & $\epsilon_{\mu\nu\lambda\sigma}p^\mu(D^+) \; \tilde{T}^{(1)\nu}(D^+)\tilde{t}^{(1)\lambda}(V_1) \; \tilde{t}^{(1)\sigma}(V_2) $ \\
$D^+[D]\to V_1V_2$ & $\tilde{T}^{(2)\mu\nu}(D^+)\tilde{t}^{(1)}_\mu(V_1)\tilde{t}^{(1)}_\nu(V_2)$ \\
$D^+\to AP_1, A[S] \to VP_2$ & $\tilde{T}^{(1)\mu}(D^+) \; P^{(1)}_{\mu\nu}(A) \; \tilde{t}^{(1)\nu}{(V)}$ \\
$D^+\to AP_1, A[D] \to VP_2$ & $\tilde{T}^{(1)\mu}(D^+) \; \tilde{t}^{(2)}_{\mu\nu}(A) \; \tilde{t}^{(1)\nu}{(V)}$ \\
$D^+\to AP_1, A\to SP_2$ & $\tilde{T}^{(1)\mu}(D^+)\tilde{t}^{(1)}_\mu{(A)}$ \\
$D^+\to VS$ & $\tilde{T}^{(1)\mu}(D^+)\tilde{t}^{(1)}_\mu{(V)}$ \\
$D^+\to V_1P_1, V_1 \to V_2P_2$ & $\epsilon_{\mu\nu\lambda\sigma}p^\mu_{V1}r^\nu_{V1}p^\lambda_{P1}r^\sigma_{V2}$ \\
$D^+\to PP_1,P\to VP_2$ & $p^\mu(P_2)\tilde{t}^{(1)}_\mu{(V)}$ \\
$D^+\to TS$ & $\tilde{T}^{(2)\mu\nu}(D^+)\tilde{t}^{(2)}_{\mu\nu}(T)$ \\
\hline
\end{tabular}
 \caption{The spin factor $S(p)$ for each decay chain. All operators, i.e.~$\tilde{t}$ and $\tilde{T}$, have the same definitions as in Ref.~\cite{covariant-tensors}. Scalar, pseudo-scalar, vector, axial-vector and tensor states are denoted by $S$, $P$, $V$, $A$ and $T$, respectively. The $[S]$, $[P]$ and $[D]$ denote the orbital angular-momentum quantum numbers $L$ = 0, 1 and 2, respectively.\label{table:spin_factors}}
\end{center}
\end{table}

\subsection{Fit results}
With the method described in Sec.~\ref{AAfitmethod}, we perform the fit in steps, by adding intermediate processes one by one. Based on previous analyses~\cite{BESIII:2017jyh,BESIII:2019ymv}, the process $D^+\to K_S^0a_1(1260)^+[S](\to \rho^+\pi^0)$ is expected to have the largest fitting fraction (FF). Hence its magnitude and phase are fixed to 1.0 and 0.0 as reference, respectively, while those of other processes are floating. The $\omega_{\rm sig}$ value in Eq.~(\ref{likelihoodfinal}) is fixed to the purity given in Sec.~\ref{AASelection}. 

Since $\bar{K}^{*0}$ and $\rho^+$ peaks are clearly observed in the corresponding invariant mass spectra, we try to add $D^+\to \bar{K}^{*0}\rho^+$ first, as well as a few processes including these two mesons. Then we test other possible intermediate resonances, including $\bar{K}_1(1400)^0$, $K_1(1270)^+$, $f_0(500)$, $f_0(980)$, $(K^+\pi^-)_{S\rm -wave}$, etc. Finally, amplitudes for $D^+\to K_S^0a_1(1260)^+[S](\to \rho^+\pi^0)$, $D^+\to K_S^0a_1(1260)^+(\to f_0(500)\pi^+)$, $D^+\to \bar{K}_1(1400)^0(\to \bar{K}^{*0}\pi^0)\pi^+$, $D^+\to \bar{K}^{*0}\rho^+$, $D^+\to \bar{K}^{*0}(\pi^+\pi^0)_V$, and $D^+\to K_S^0(\rho^+\pi^0)_P$, which have statistical significance greater than 5 standard deviations, are retained in the nominal fit. The statistical significance of each process is determined from the changes in log-likelihood and the numbers of degrees of freedom when the fits are performed with and without the process included.

Generator-level MC events without detector acceptance and resolution effects are used to calculate the FFs for individual amplitudes. The FF for the $n^{\rm th}$ amplitude is defined as
\begin{eqnarray}\begin{aligned}
	{\rm FF}_{n} = \frac{\sum^{N_{\rm gen}} \left|c_n\mathcal A_{n}\right|^{2}}{\sum^{N_{\rm gen}} \left|\mathcal M\right|^{2}}\,, \label{Fit-Fraction-Definition}
\end{aligned}\end{eqnarray}
where $N_{\rm gen}$ is the number of phase space MC events at the generator level. The sum of these FFs is generally not unity due to net constructive or destructive interference. Interference~(IN) between the $n^{\rm th}$ and $n^{\rm \prime th}$ amplitudes is defined as
\begin{equation}
    {\rm IN}_{nn^\prime}
    = \frac{\sum^{N_{\rm gen}}2{\rm Re}[c_nc^*_{n^\prime}\mathcal{A}_n\mathcal{A}^{*}_{n^\prime}]}{\sum^{N_{\rm gen}}|\mathcal{M}|^2}.
\label{inter}
\end{equation}
The statistical uncertainties of the FFs are obtained by randomly perturbing the fit parameters according to their uncertainties and the covariance matrix and re-evaluating FFs. A Gaussian function is used to fit the resulting distribution for each FF and the fitted width is taken as its statistical uncertainty.  

\begin{figure}[htbp]
          \centering
          \includegraphics[width=6.7cm]{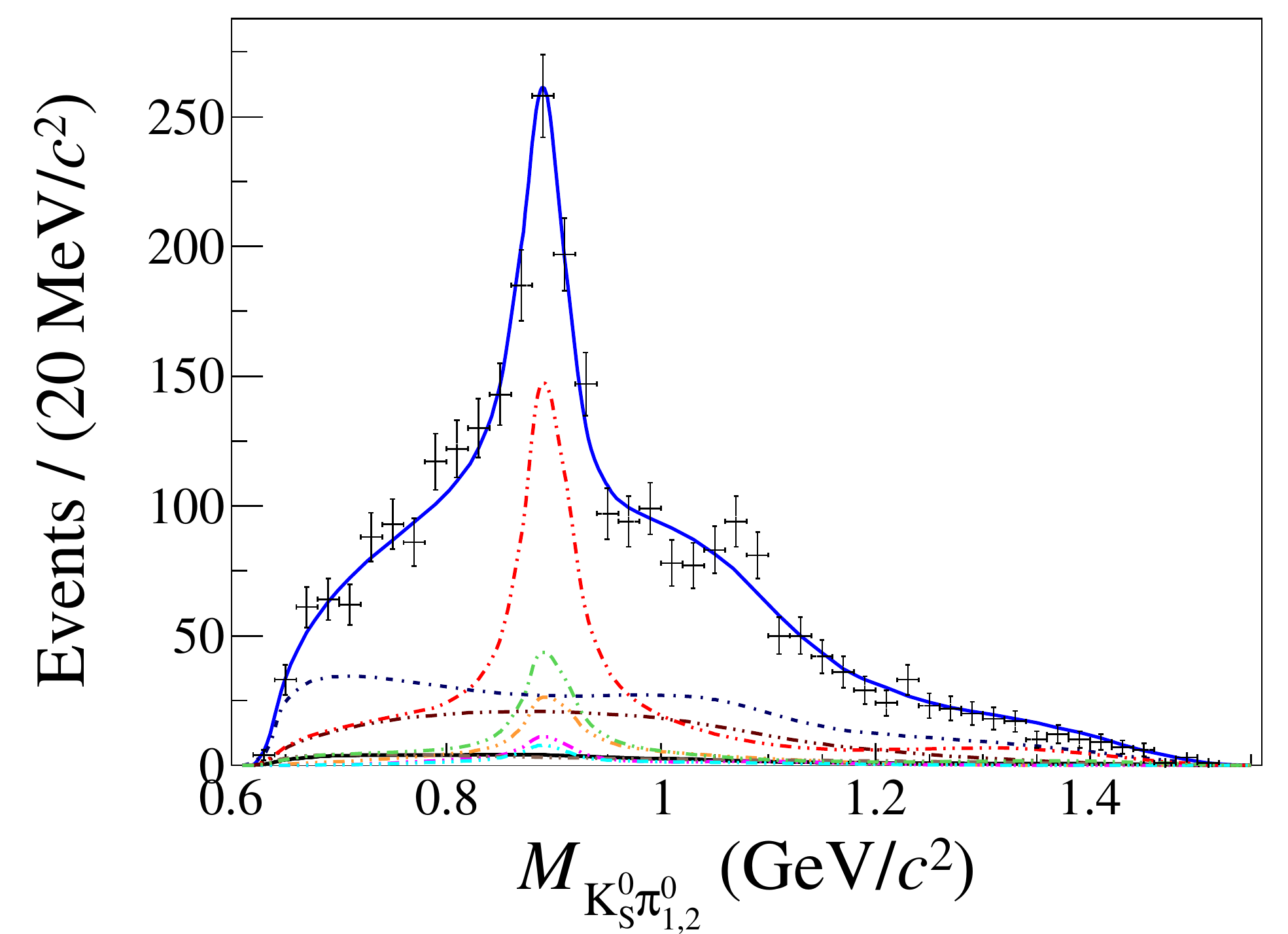}
          \includegraphics[width=6.7cm]{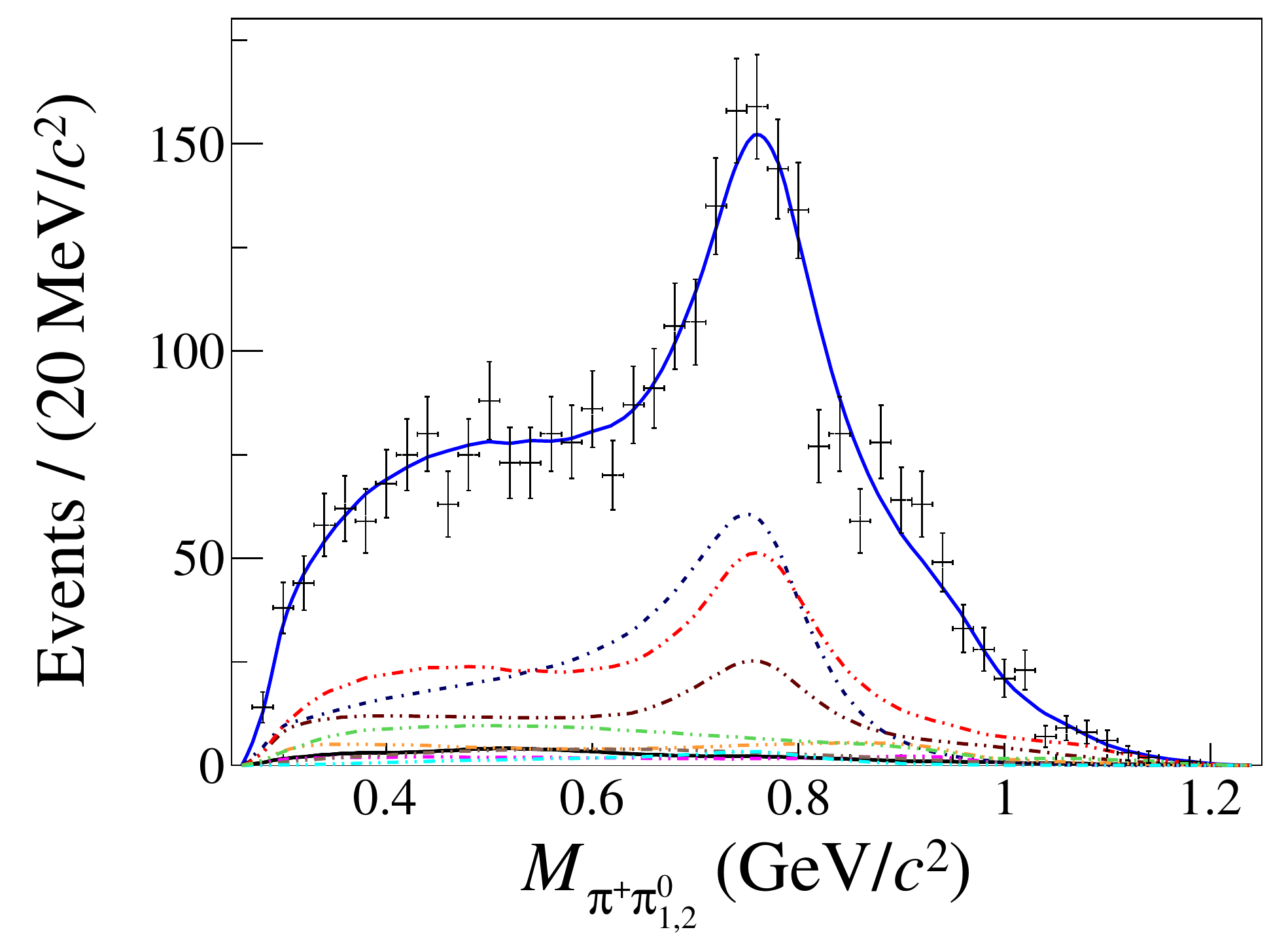}\\
          \includegraphics[width=6.7cm]{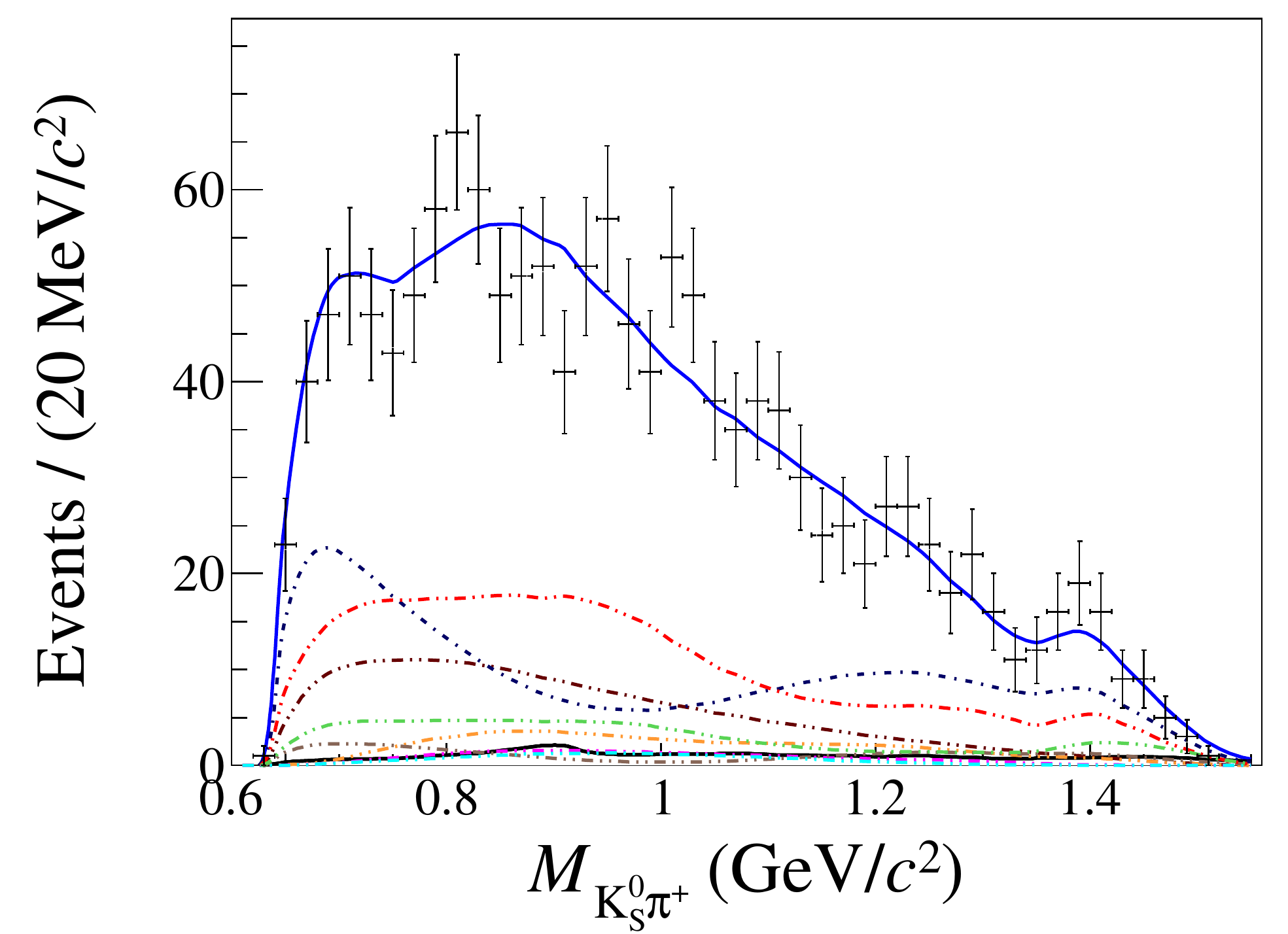}
          \includegraphics[width=6.7cm]{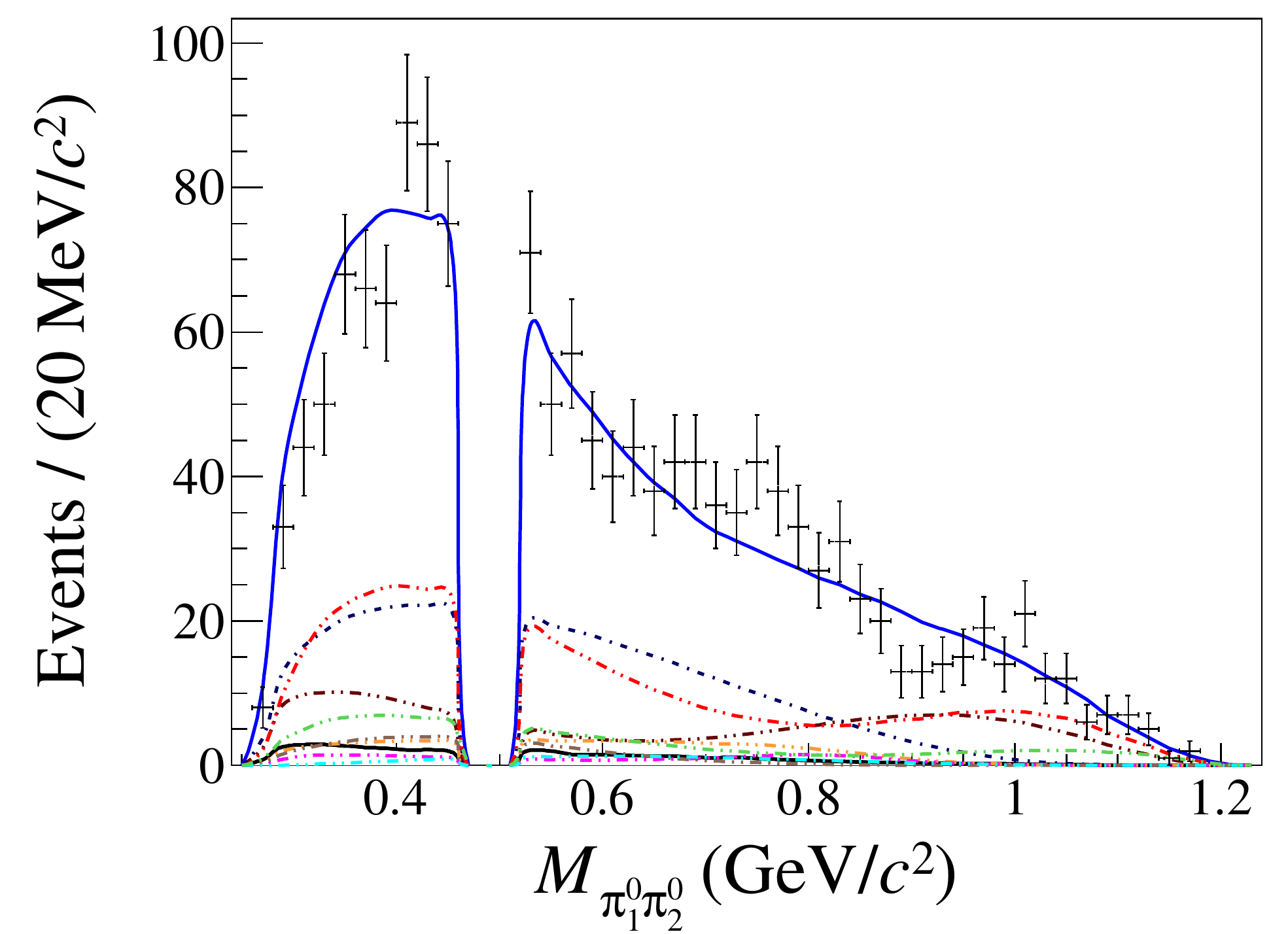}\\
          \includegraphics[width=6.7cm]{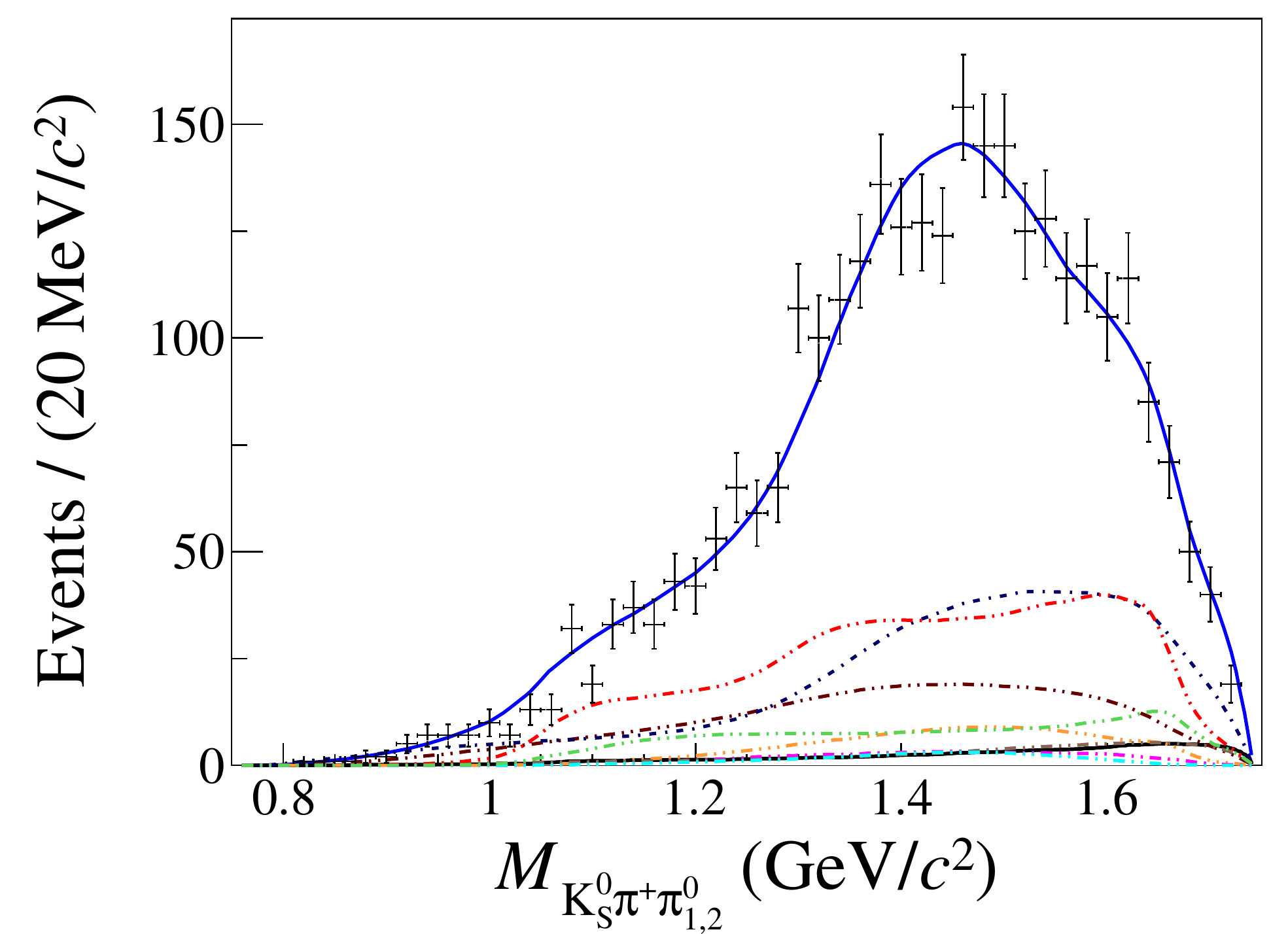}
          \includegraphics[width=6.7cm]{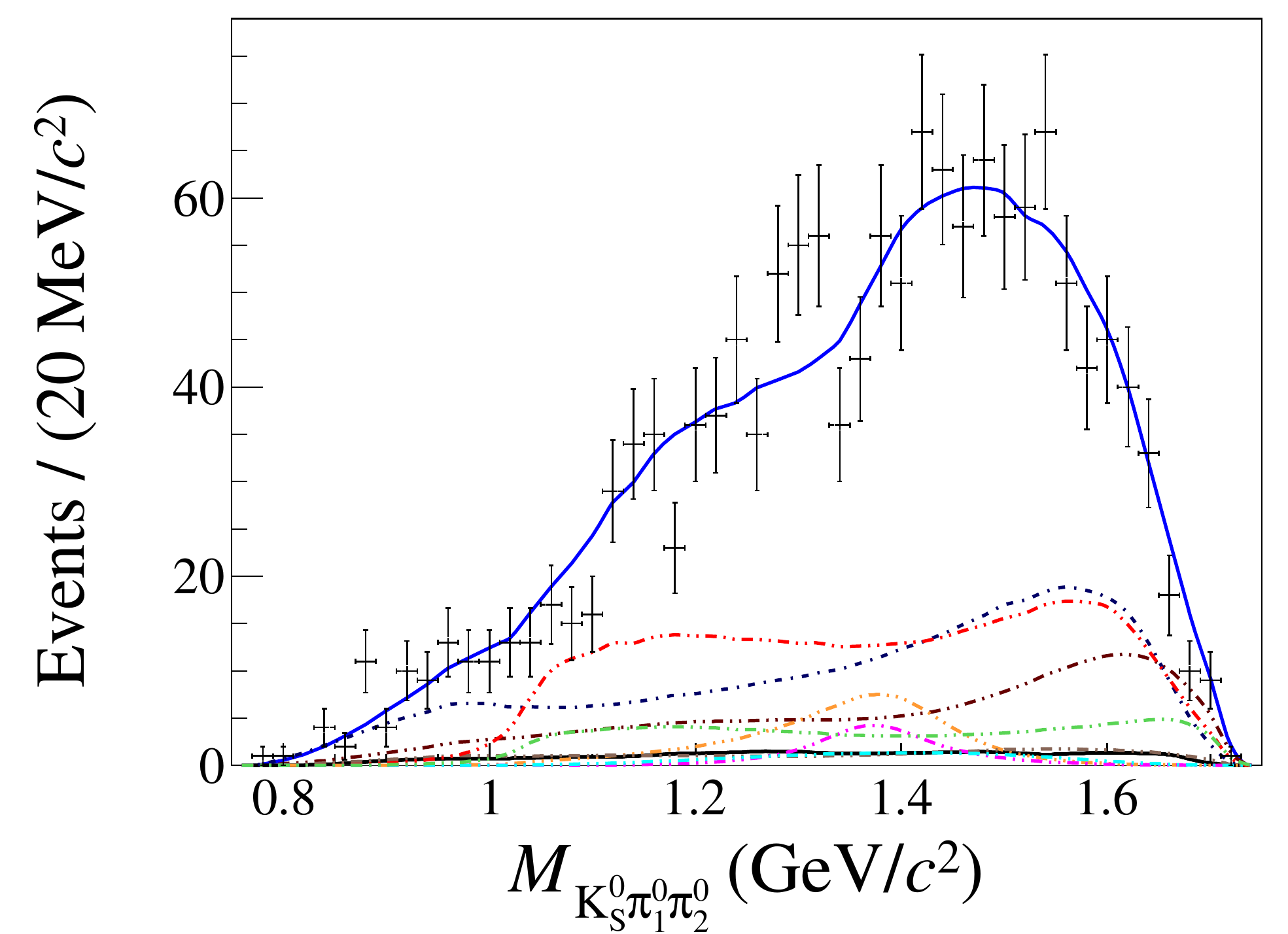}\\
          \includegraphics[width=6.7cm]{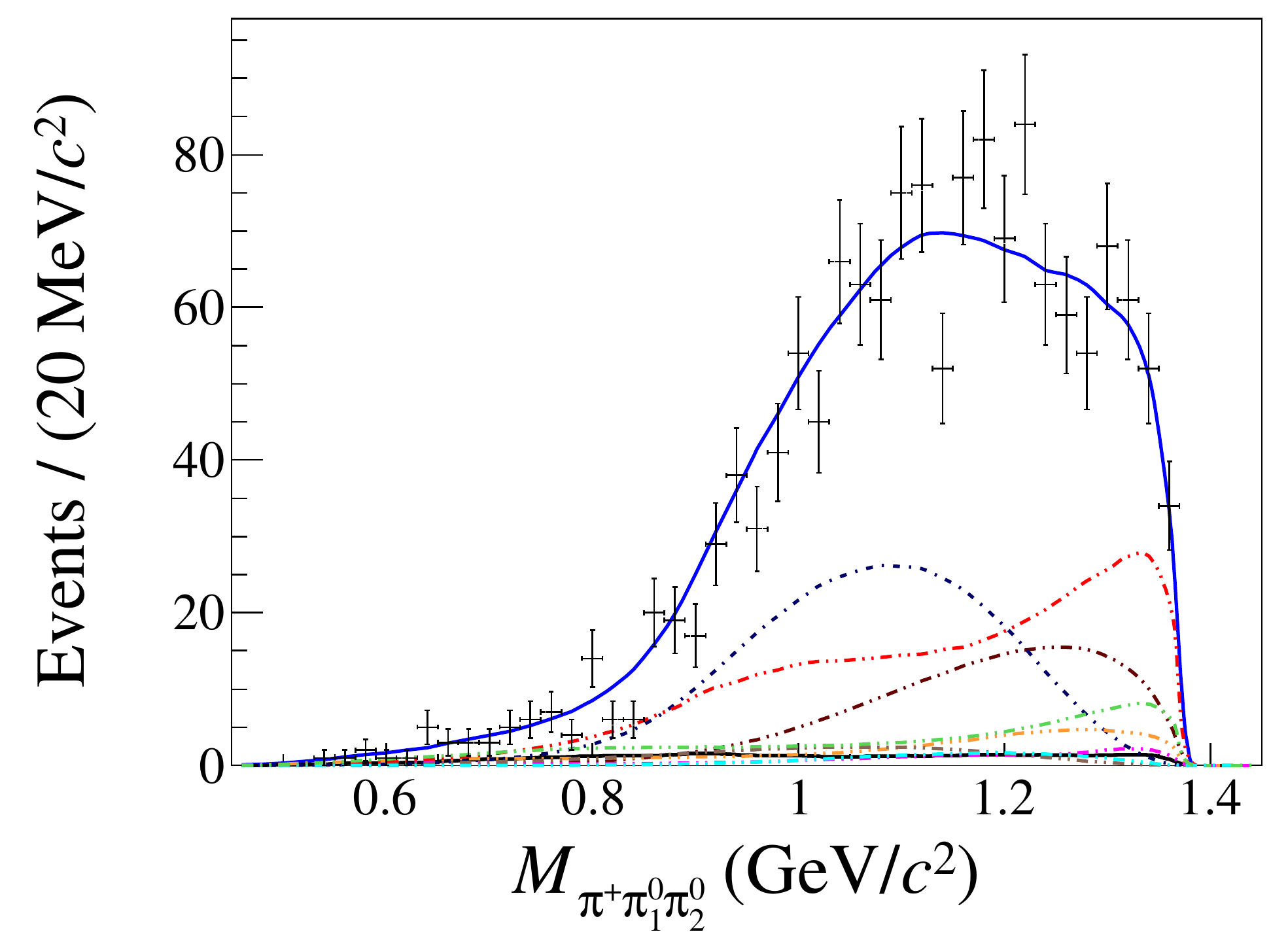}
          \includegraphics[width=6.7cm]{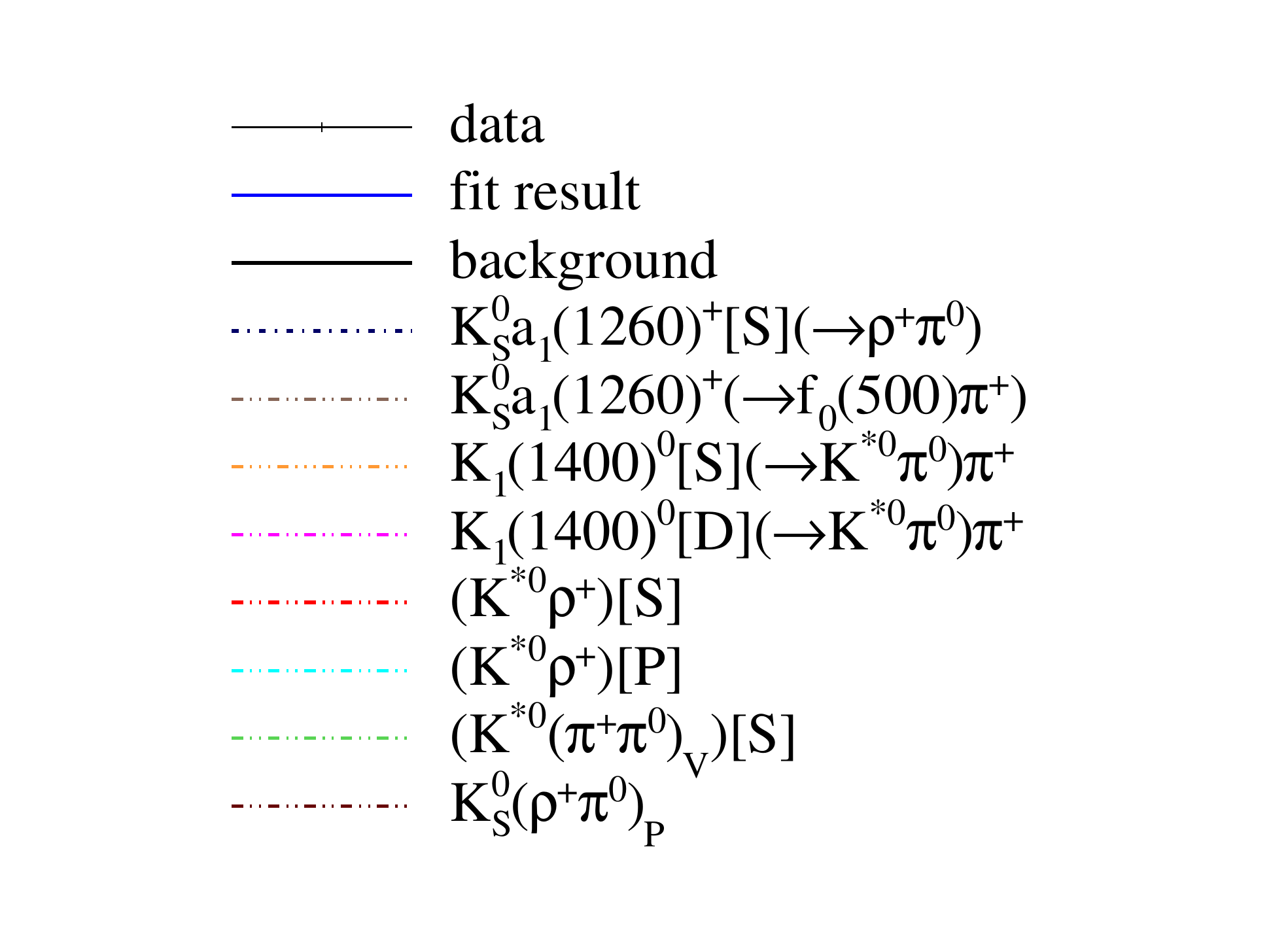}\\
          \caption{Projections of the nominal fit onto invariant mass distributions. The data sample is represented by points with error bars, the fit results by the solid blue lines, and the background by the solid black lines. Colored curves show the components of the fit model. The two $\pi^0$ are sorted according to the magnitude of their momentum. The labels $\pi^0_{1,2}$ mean that two distributions involving a single $\pi^0$ have been combined.}
    \label{pwa:proji}
\end{figure}

According to the fit result, the phases, FFs and statistical significances for various amplitudes are listed in table~\ref{tab:signi}. The interference between processes is listed in table~\ref{tab:inter} of appendix~\ref{app:inter}.
The statistical significances for the processes tested but not included in the nominal fit are listed in appendix~\ref{app:others}. The mass projections of the nominal fit are shown in figure~\ref{pwa:proji}.  

\begin{table}[htbp]
	\centering
 \resizebox{\textwidth}{!}{
	\begin{tabular}{| l r@{ $\pm$ }c@{ $\pm$ }c r@{ $\pm$ }c@{ $\pm$ }c c |}
	\hline
  Amplitude &\multicolumn{3}{c}{Phase $\phi_n$ (rad)} &\multicolumn{3}{c}{FF (\%)} &Significance ($\sigma$)\\
	\hline
    $D^+\to K_S^0a_1(1260)^+[S](\to \rho^+\pi^0)$
    &\multicolumn{3}{c}{{0.0 (fixed)}} &30.0 &3.6 &4.2 &$>$10\\
    \hline
    $D^+\to K_S^0a_1(1260)^+(\to f_0(500)\pi^+)$
    & 4.78 &0.22 &0.20  &3.5  &1.1 &1.9  &6.9\\
    \hline
    $D^+\to \bar{K}_1(1400)^0[S](\to \bar{K}^{*0}\pi^0)\pi^+$
    &-3.01 &0.12 &0.16  &6.0  &1.2 &0.3  &9.6 \\
    $D^+\to \bar{K}_1(1400)^0[D](\to \bar{K}^{*0}\pi^0)\pi^+$
    & 4.29 &0.16 &0.20  &2.4  &0.6 &0.2  &6.7 \\
    $D^+\to \bar{K}_1(1400)^0(\to \bar{K}^{*0}\pi^0)\pi^+$
    &\multicolumn{3}{c}{-}           &8.0  &1.2 &0.4  &- \\
    \hline
    $D^+[S]\to \bar{K}^{*0}\rho^+$
    &-3.33 &0.10 &0.17  &31.8 &2.7 &1.3  &$>$10\\
    $D^+[P]\to \bar{K}^{*0}\rho^+$
    &-1.68 &0.17 &0.16  &1.7  &0.6 &0.1  &5.0\\
    $D^+\to \bar{K}^{*0}\rho^+$
    &\multicolumn{3}{c}{-}     &33.6 &2.7 &1.4  &- \\
    \hline
    $D^+[S]\to \bar{K}^{*0}(\pi^+\pi^0)_V$
    &-5.60 &0.13 &0.16  &9.1  &2.0 &1.0  &9.4 \\ \hline
    $D^+\to K_S^0(\rho^+\pi^0)_P$
    & 0.76 &0.11 &0.24  &16.5 &1.6 &0.3  &$>$10\\
    \hline
    \end{tabular}
    }
	\caption{The phases, FFs and statistical significances for various amplitudes in the nominal fit. Groups of related amplitudes are separated by horizontal lines. The last row of each group gives the total fitting fractions of the above components with interference considered. The first and second uncertainties of the phases and FFs are statistical and systematic, respectively. The $\bar{K}^{*0}$ resonance decays to $K_S^0\pi^0$. The $\rho^+$ resonance decays to $\pi^+\pi^0$. The $f_0(500)$ resonance decays to $\pi^0\pi^0$. The total FF is {{100.39}}\%.}
	\label{tab:signi}
\end{table}

\subsection{Systematic uncertainties for the amplitude analysis}
\label{sec:PWA-Sys}
The systematic uncertainties for the amplitude analysis are summarized in table~\ref{pwa:sys}, and are described below. The square roots of the quadratic sums of each uncertainty are considered as the total uncertainties.
\begin{itemize}
\item[\lowercase\expandafter{\romannumeral1}]
Fixed parameters in the amplitudes. The masses and widths of $\bar{K}^{*0}$, $\rho^+$, $a_1(1260)^+$ and $\bar{K}_1(1400)^0$ are varied by their uncertainties~\cite{PDG}. The uncertainties of the lineshape for the $f_0(500)$ are estimated by replacing the  propagator with the RBW formula, in which the mass and width for the $f_0(500)$ are fixed at 526~MeV/$c^2$ and 534~MeV, respectively~\cite{Pelaez:2015qba}. Since varying the propagator results in different normalization factors, only the effect on all FFs is considered. The changes of the phases $\phi$ and FFs are assigned as the associated systematic uncertainties.

\item[\lowercase\expandafter{\romannumeral2}]
$R$ values. The estimation of the systematic uncertainty associated with the $R$ parameters in the Blatt-Weisskopf factors is performed by repeating the fit procedure after varying the effective radius of the intermediate states and $D^+$ meson by $\pm 1$ GeV$^{-1}$.

\item[\lowercase\expandafter{\romannumeral3}]
Fit bias. An ensemble of 600 signal MC samples is generated according to the result of the amplitude analysis. The pull distributions, supposed to be normal distributions, are used to validate the fit performance and are fitted with a Gaussian function. The fitted mean values for FFs of $D^+\to \bar{K}_1(1400)^0[S](\to \bar{K}^{*0}\pi^0)\pi^+$ deviate upward from zero by more than three standard deviations. No significant deviations are observed for other terms. We correct all FFs and phases by the fitted mean values, and assign the uncertainties of the fitted mean values as the systematic uncertainties.

\item[\lowercase\expandafter{\romannumeral4}]
Background estimation. The uncertainty from the size of the background is studied by varying the signal fraction (equivalent to the fraction of background), i.e.~$\omega_{\rm sig}$ in Eq.~(\ref{eq:likefinal}), within its corresponding statistical uncertainty. Another source of uncertainty is  the simulation of the background shape. We extract the shape with other input variables and change the fraction of different background components in MC simulation.

\item[\lowercase\expandafter{\romannumeral5}]
Experimental effects. The systematic uncertainty from the $\gamma_{\epsilon}$ factor in Eq.~(\ref{likelihood3}), which corrects for data-MC differences in tracking, PID as well as $\pi^0$ and $K_S^0$ reconstruction efficiencies, is evaluated by performing the fit after varying the weights according to their uncertainties.  
\end{itemize}

\begin{table}[h]\small
\renewcommand\arraystretch{1.25}
  \centering
	\begin{tabular}{| lccccccc |}
    \hline
    \multirow{2}{*}{Amplitude}&\multicolumn{7}{c}{Source}\vline\cr
		& & \lowercase\expandafter{\romannumeral1} &\lowercase\expandafter{\romannumeral2} &\lowercase\expandafter{\romannumeral3} &\lowercase\expandafter{\romannumeral4} &\lowercase\expandafter{\romannumeral5} 
		& Total   \\
	\hline
    $D^+\to K_S^0a_1(1260)^+[S](\to \rho^+\pi^0)$
    &FF     &1.19 &0.40 &0.04 &0.34 &0.04 &1.30 \\
    \hline
    \multirow{2}{*}{$D^+\to K_S^0a_1(1260)^+(\to f_0(500)\pi^+$)}
    &$\phi$ &0.91 &0.56 &0.04 &0.10 &0.05 &1.08 \\
    &FF     &1.69 &0.08 &0.04 &0.56 &0.02 &1.78 \\
    \hline
    \multirow{2}{*}{$D^+\to \bar{K}_1(1400)^0[S](\to \bar{K}^{*0}\pi^0)\pi^+$}
    &$\phi$ &1.32 &0.16 &0.04 &0.01 &0.05 &1.33 \\
    &FF     &0.24 &0.07 &0.04 &0.23 &0.01 &0.34 \\
    \multirow{2}{*}{$D^+\to \bar{K}_1(1400)^0[D](\to \bar{K}^{*0}\pi^0)\pi^+$}
    &$\phi$ &1.24 &0.14 &0.04 &0.23 &0.02 &1.27 \\
    &FF     &0.39 &0.04 &0.04 &0.18 &0.01 &0.43 \\
    $D^+\to \bar{K}_1(1400)^0(\to \bar{K}^{*0}\pi^0)\pi^+$
    &FF     &0.33 &0.10 &0.04 &0.29 &0.01 &0.45 \\
    \hline
    \multirow{2}{*}{$D^+[S]\to \bar{K}^{*0}\rho^+$}
    &$\phi$ &1.67 &0.17 &0.04 &0.02 &0.06 &1.68 \\
    &FF     &0.50 &0.32 &0.04 &0.03 &0.02 &0.60 \\
    \multirow{2}{*}{$D^+[P]\to \bar{K}^{*0}\rho^+$}
    &$\phi$ &0.94 &0.09 &0.04 &0.07 &0.01 &0.95 \\
    &FF     &0.16 &0.04 &0.04 &0.14 &0.00 &0.22 \\
    $D^+\to \bar{K}^{*0}\rho^+$
    &FF     &0.52 &0.32 &0.04 &0.00 &0.02 &0.61 \\
    \hline
    \multirow{2}{*}{$D^+[S]\to \bar{K}^{*0}(\pi^+\pi^0)_V$}
    &$\phi$ &1.20 &0.06 &0.04 &0.03 &0.05 &1.20 \\
    &FF     &0.51 &0.22 &0.04 &0.11 &0.02 &0.57 \\
    \hline
    \multirow{2}{*}{$D^+\to K_S^0(\rho^+\pi^0)_P$}
    &$\phi$ &2.20 &0.58 &0.05 &0.38 &0.10 &2.31 \\
    &FF     &0.17 &0.32 &0.04 &0.02 &0.02 &0.37 \\
	\hline
	\end{tabular}
	  \caption{Systematic uncertainties on the $\phi$ and FF for each amplitude in units of the corresponding statistical uncertainty. The sources are:
    (\lowercase\expandafter{\romannumeral1}) fixed parameters in the amplitudes,
    (\lowercase\expandafter{\romannumeral2}) $R$ values,
    (\lowercase\expandafter{\romannumeral3}) fit bias,
	(\lowercase\expandafter{\romannumeral4}) background estimation,
	(\lowercase\expandafter{\romannumeral5}) experimental effects.
}
	\label{pwa:sys}
\end{table}

\section{BF measurement}
\label{BFSelection}
The BF measurement is based on the following equations:
\begin{equation}
    N^{\rm ST}_{\rm tag}=2N_{D^+D^-}\mathcal{B}_{\rm tag}\epsilon^{\rm ST}_{\rm tag},
     \label{eq-ST}
\end{equation}
\begin{equation}
N_{\text{tag,sig}}^{\text{DT}}=2N_{D^{+}D^{-}}\mathcal{B}_{\text{tag}}\mathcal{B}_{\text{sig}}\epsilon_{\text{tag,sig}}^{\text{DT}}\,,
\end{equation}
where $N_{D^+D^-}$ is the total number of $D^+D^-$ pairs produced in the initial $e^+e^-$ collisions; $N^{\rm ST}_{\rm tag}$ is the ST yield for a specific tag mode; $N^{\rm DT}_{\rm tag,sig}$ is the DT yield; $\mathcal{B}_{\rm tag}$  and $\mathcal{B}_{\rm sig}$ are the BFs of the tag and the signal modes, respectively; $\epsilon^{\rm ST}_{\rm tag}$ is the ST efficiency to reconstruct the tag mode; $\epsilon^{\rm DT}_{\rm tag,sig}$ is the DT efficiency to reconstruct both the tag and the signal decay modes. The total DT yield is calculated as
\begin{equation}
  N^{\rm DT}_{\rm total} = \sum_{\alpha}{N^{\rm DT}_{\alpha,\rm sig}} = \mathcal{B}_{\rm sig}\sum_{\alpha}{2N_{D^+D^-}\mathcal{B}_{\alpha}\epsilon^{\rm DT}_{\alpha,\rm sig}},
\end{equation}
where $\alpha$ represents different tag modes. By isolating $\mathcal{B}_{\rm sig}$, we obtain:
\begin{equation}
  \mathcal{B}_{\rm sig}=\frac{N^{\rm DT}_{\rm total}}{\mathcal{B_{\rm sub}}\sum_{\alpha}{N^{\rm ST}_{\alpha}}\epsilon^{\rm DT}_{\alpha,\rm sig}/\epsilon^{\rm ST}_{\alpha}},
  \label{abs:bf}
\end{equation}
where $\mathcal{B_{\rm sub}}=\mathcal{B}_{K_S^0\to\pi^+\pi^-}\mathcal{B}^2_{\pi^0\to\gamma\gamma}$ is introduced to take into account the fact that the signal is reconstructed through these decays. The yields $N^{\rm DT}_{\rm total}$ and $N^{\rm ST}_{\alpha}$ are obtained from the data sample, while $\epsilon^{\rm ST}_{\alpha}$ and $\epsilon^{\rm DT}_{\alpha,\rm sig}$ can be obtained from the inclusive and signal MC samples in which $D^{+} \to K_S^0\pi^{+}\pi^{0}\pi^0$ events are generated according to the result of the amplitude analysis, respectively.

\begin{figure*}[htp]
\begin{center}
    \includegraphics[width=6.cm]{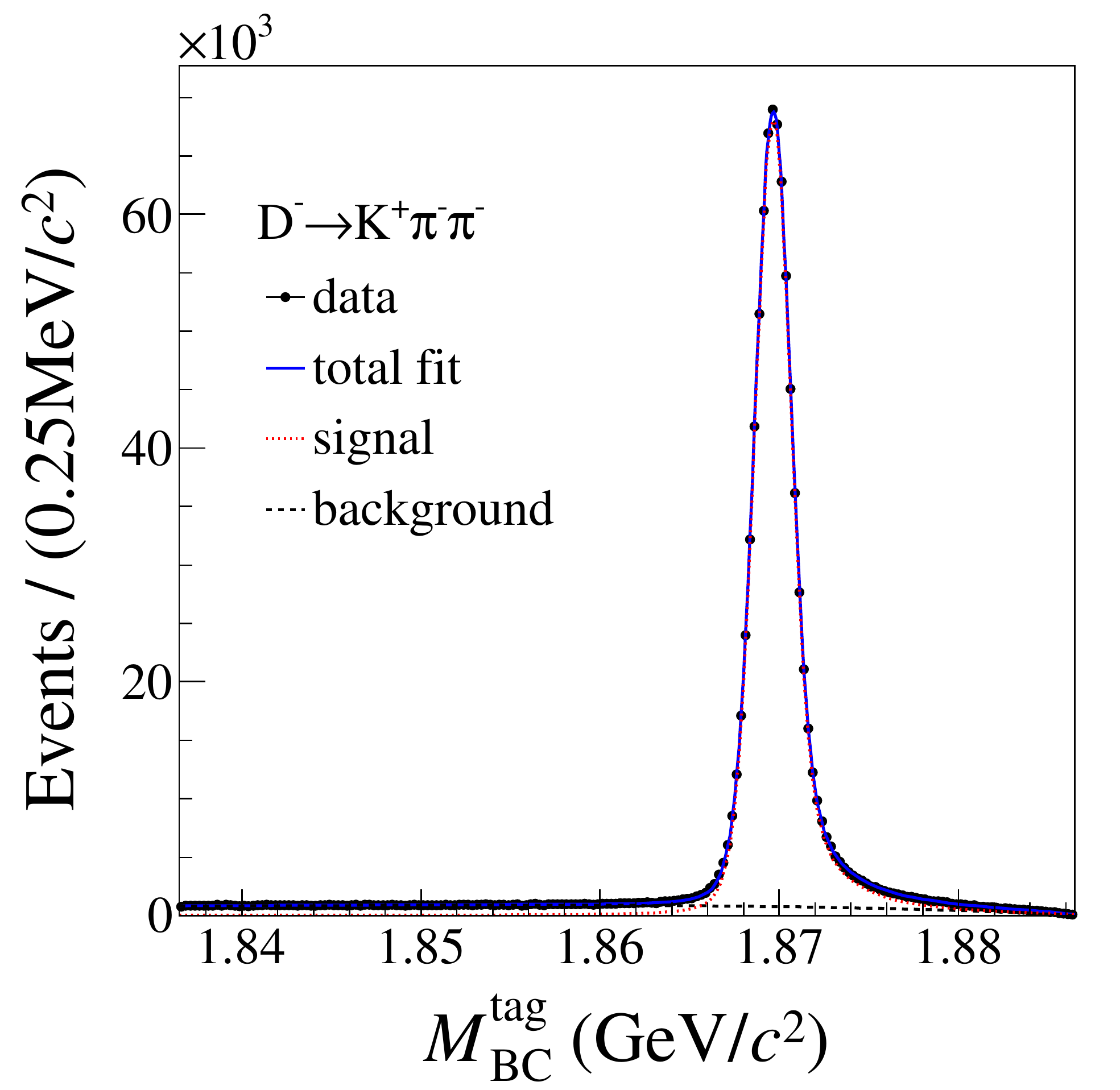}
    \includegraphics[width=6.cm]{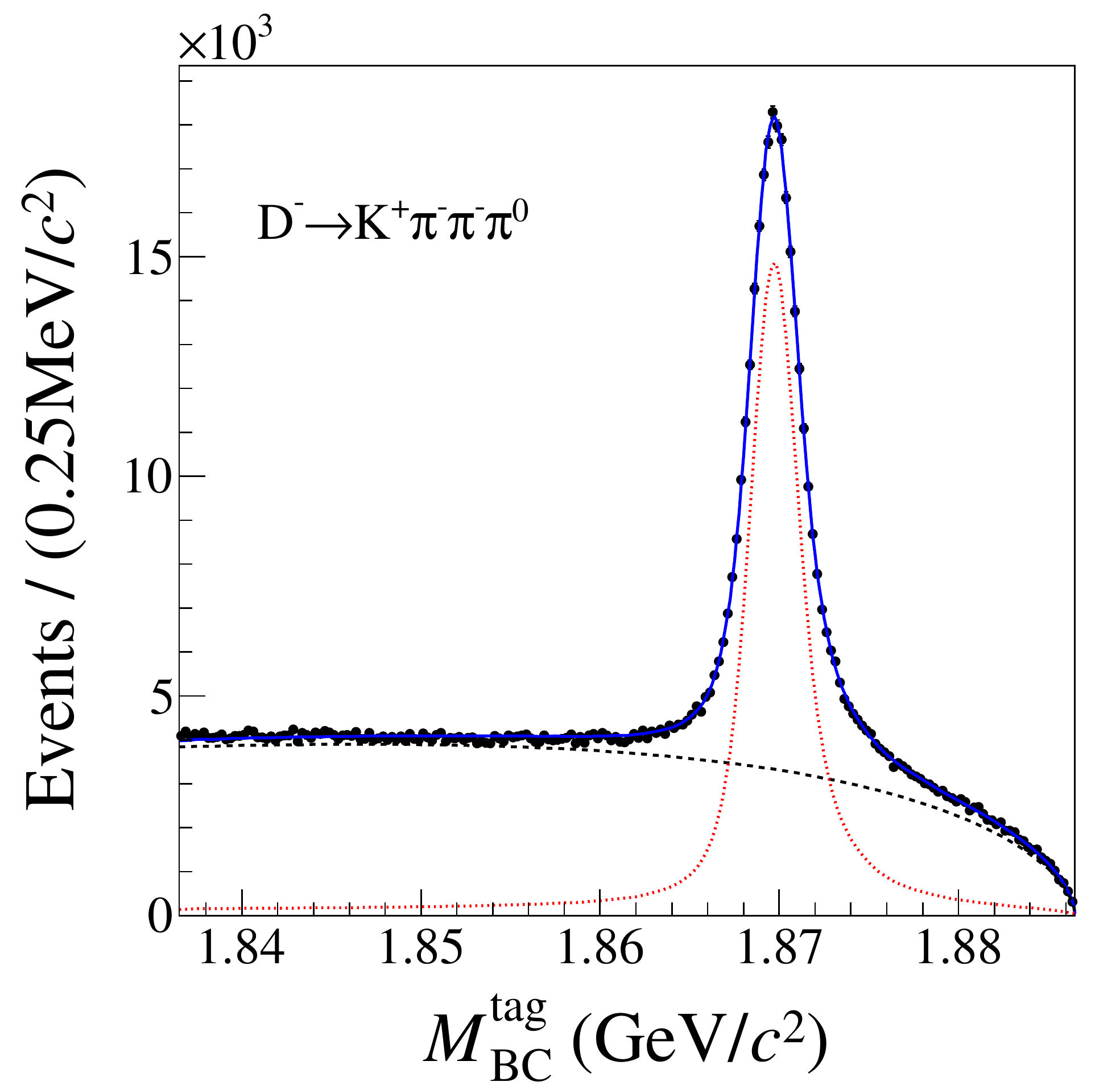}
    \includegraphics[width=6.cm]{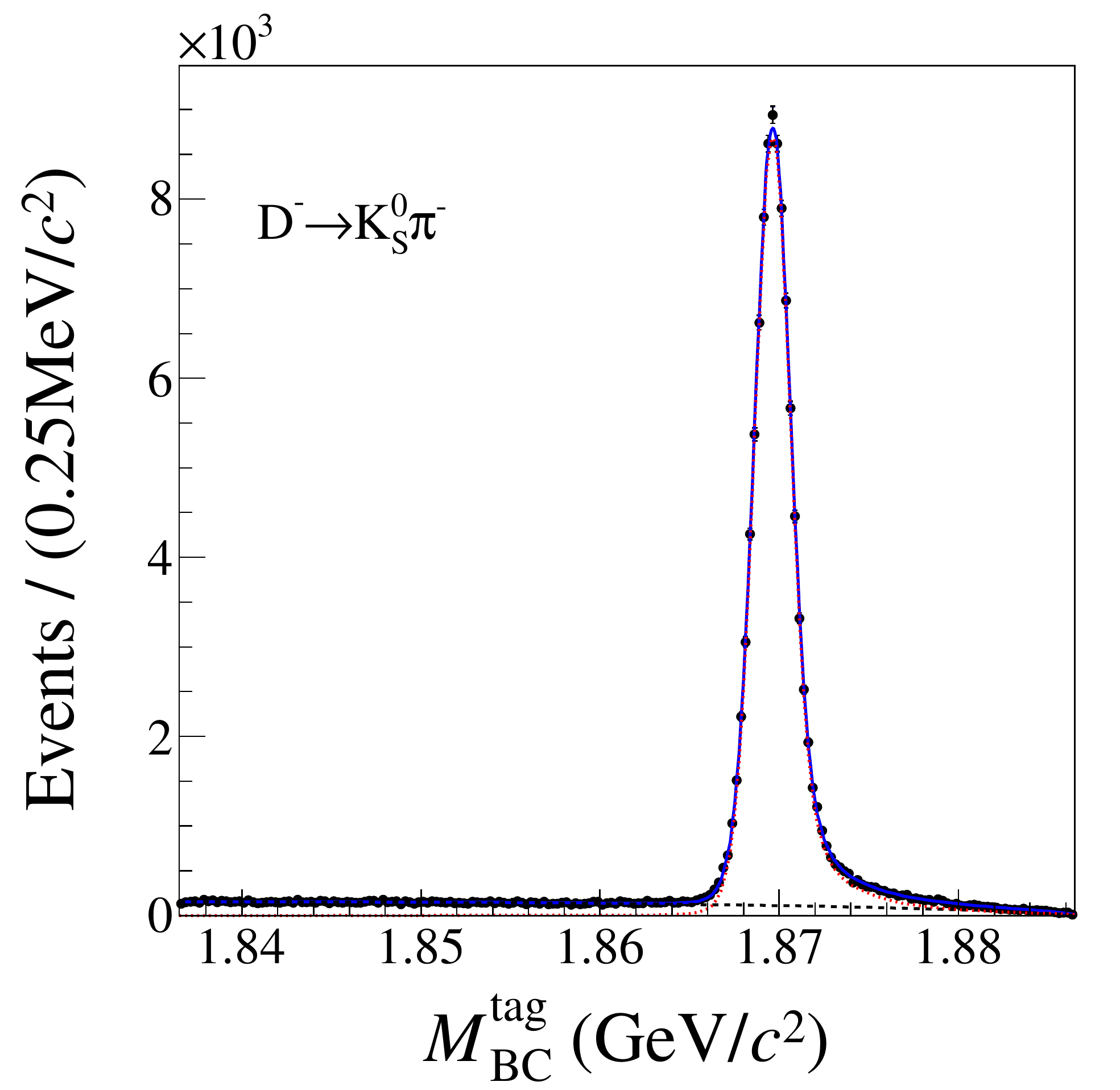}
    \includegraphics[width=6.cm]{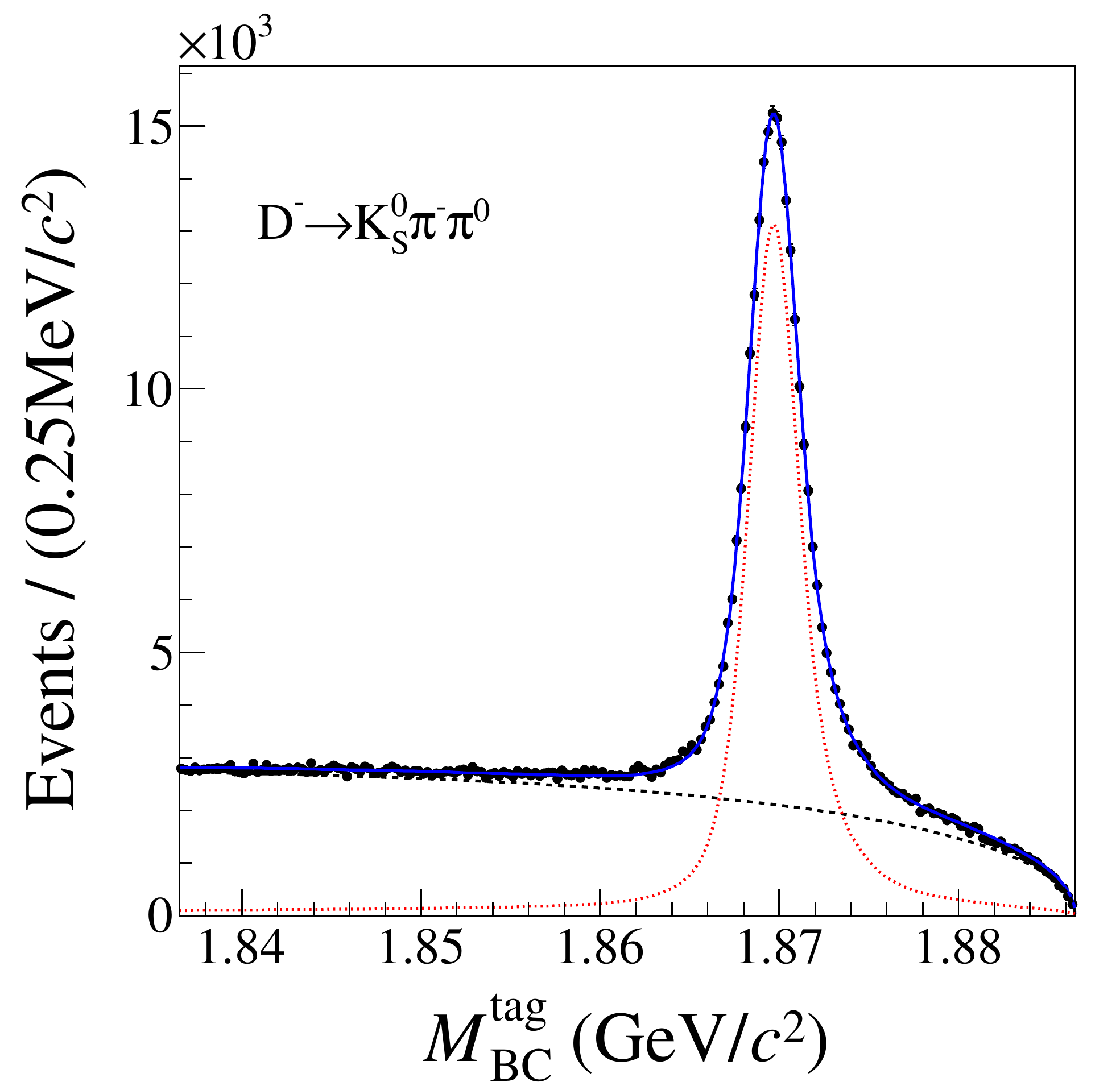}
    \includegraphics[width=6.cm]{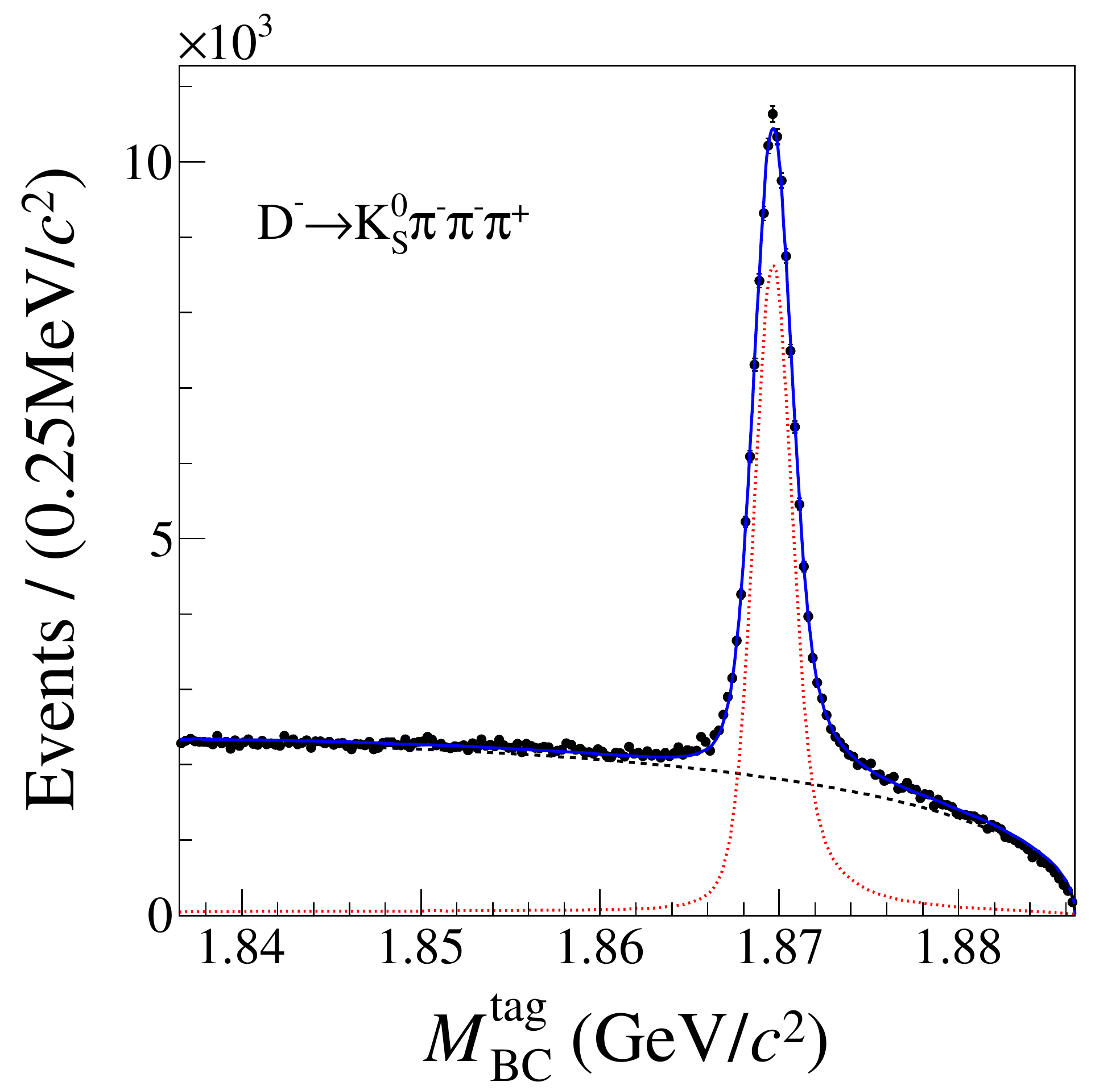}
    \includegraphics[width=6.cm]{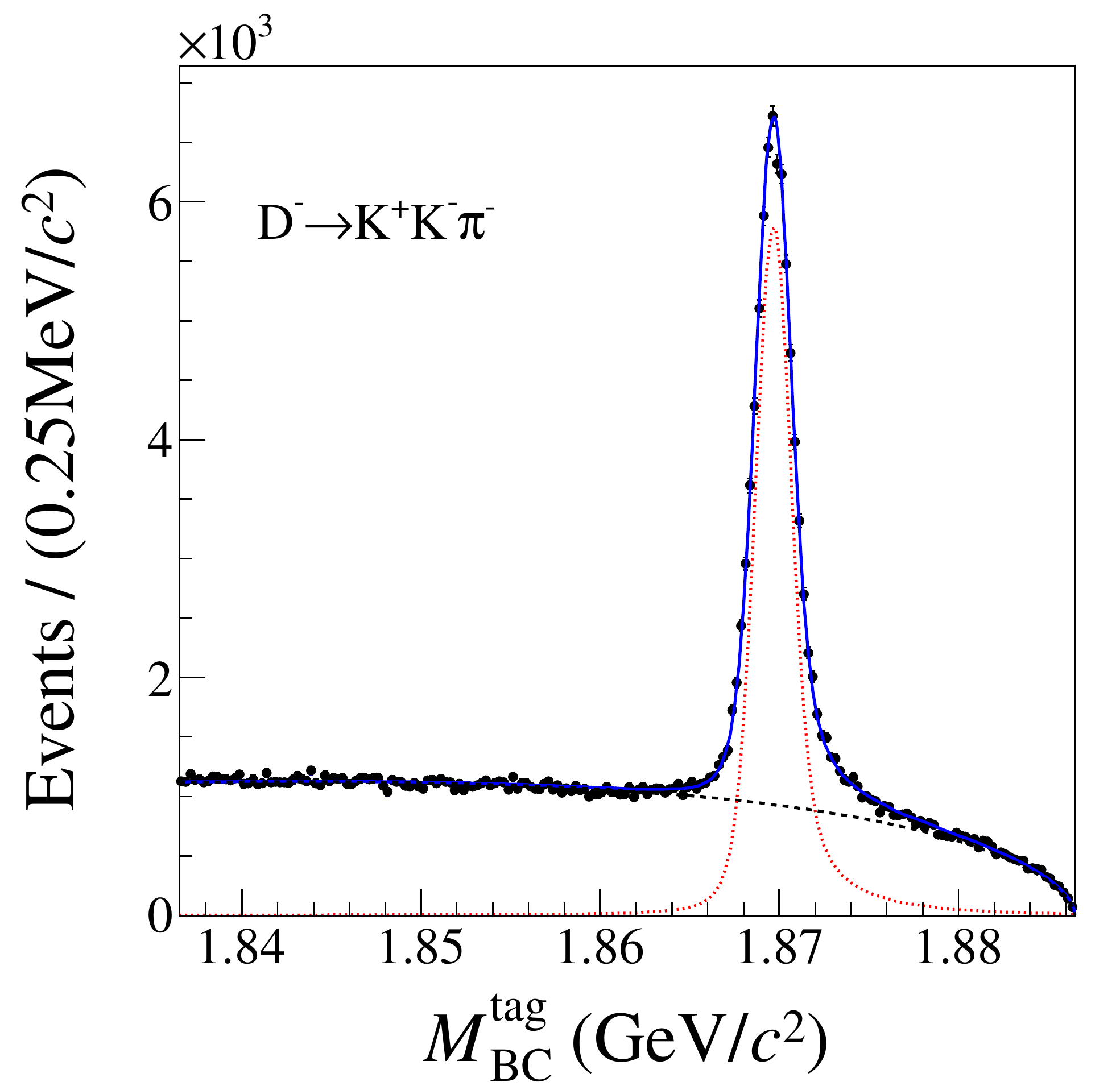}
\caption{Fits to the $M^{\rm tag}_{\rm BC}$ distribution of the ST candidates. The dots with error bars are data, the solid blue lines are the total fit, the black dashed lines represent the background shape, the red dashed lines represent the signal shape.}
\label{fig:ST_yield}
\end{center}
\end{figure*}

Six tag modes used in the BF measurement and their energy difference requirements are listed in table~\ref{tab:ST_yield}. For multiple ST candidates, the one with minimum $|\Delta{E}|$ is chosen. The ST yields~($N_{\rm tag}^{\rm ST}$) and efficiencies ($\epsilon_{\rm tag}^{\rm ST})$ for each tag mode, also listed in table~\ref{tab:ST_yield}, are obtained by fitting the corresponding $M_{\rm BC}^{\rm tag}$ distributions individually. In the fit, the signal is modeled by a MC-simulated shape convolved with a Gaussian function which describes the resolution difference between data and MC simulation. The background is described by the ARGUS~\cite{ARGUS:1990hfq} function whose parameters are left floating except for the endpoint, which is fixed at 1.8865~GeV. Figure~\ref{fig:ST_yield} shows the fit results.

\begin{table}[hbtp]
  \begin{center}
   \resizebox{\textwidth}{!}{
    \begin{tabular}{|l c r@{ $\pm$ }l r@{ $\pm$ }l r@{ $\pm$ }l r@{ $\pm$ }l|}
      \hline
      Tag mode &$\Delta E$ (MeV) & \multicolumn{2}{c}{$N_{\rm tag}^{\rm ST}$}  &\multicolumn{2}{c}{$\epsilon_{\rm tag}^{\rm ST}$ $(\%)$}  &\multicolumn{2}{c}{$\epsilon_{\rm tag,sig}^{\rm DT}$ $(\%)$} &\multicolumn{2}{c}{$\epsilon_{\rm sig}$ $(\%)$}\vline\\
      \hline
      $D^-\to K^+\pi^-\pi^-$        &(-25, 25) &821313 &974 &52.65 &0.02 &6.46 &0.01 &12.27 &0.02 \\
      $D^-\to K^+\pi^-\pi^-\pi^0$   &(-55, 40) &285779 &855 &29.37 &0.03 &2.99 &0.01 &10.19 &0.03 \\
      $D^-\to K_S^0\pi^-$           &(-25, 25) &101444 &339 &55.38 &0.07 &6.71 &0.03 &12.12 &0.05 \\
      $D^-\to K_S^0\pi^-\pi^0$      &(-55, 40) &249765 &766 &30.72 &0.03 &3.25 &0.01 &10.58 &0.03 \\
      $D^-\to K_S^0\pi^-\pi^-\pi^+$ &(-25, 25) &119226 &493 &30.02 &0.04 &3.46 &0.01 &11.53 &0.04 \\
      $D^-\to K^+K^-\pi^-$          &(-25, 25) &70825  &337 &42.75 &0.07 &5.18 &0.02 &12.11 &0.06 \\
      \hline
    \end{tabular}
   }
  \end{center}
      \caption{The $\Delta E$ requirements, ST yields ($N_{\rm tag}^{\rm ST}$), ST efficiencies ($\epsilon_{\rm tag}^{\rm ST}$), DT efficiencies ($\epsilon_{\rm tag,sig}^{\rm DT})$ and signal efficiencies ($\epsilon_{\rm sig}=\epsilon_{\rm tag,sig}^{\rm DT}/\epsilon_{\rm tag}^{\rm ST}$) for six tag modes. The uncertainties are statistical only.}
       \label{tab:ST_yield}
\end{table}

\begin{figure*}[htp]
\begin{center}
    \includegraphics[width=6.cm]{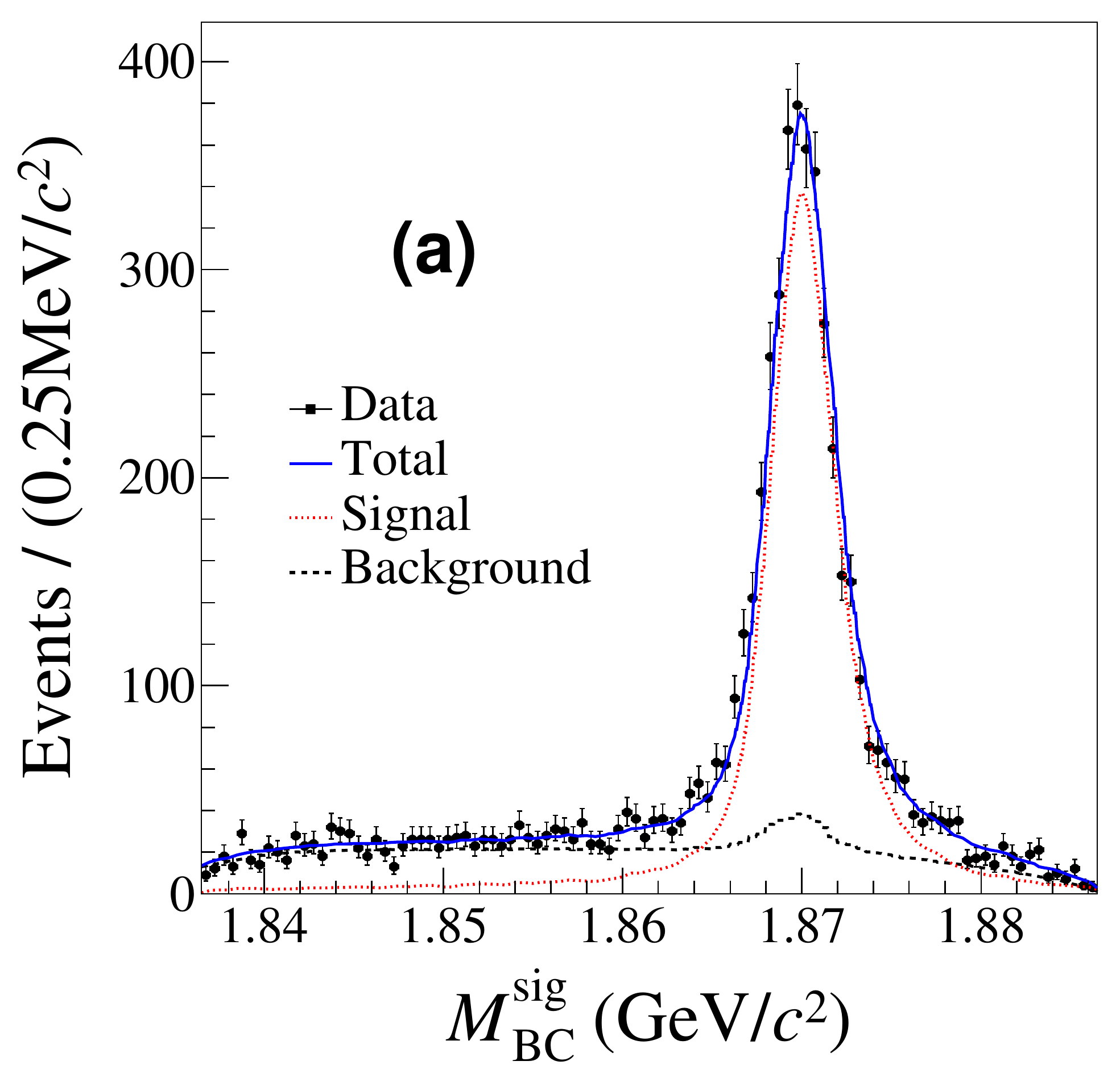}
    \includegraphics[width=6.cm]{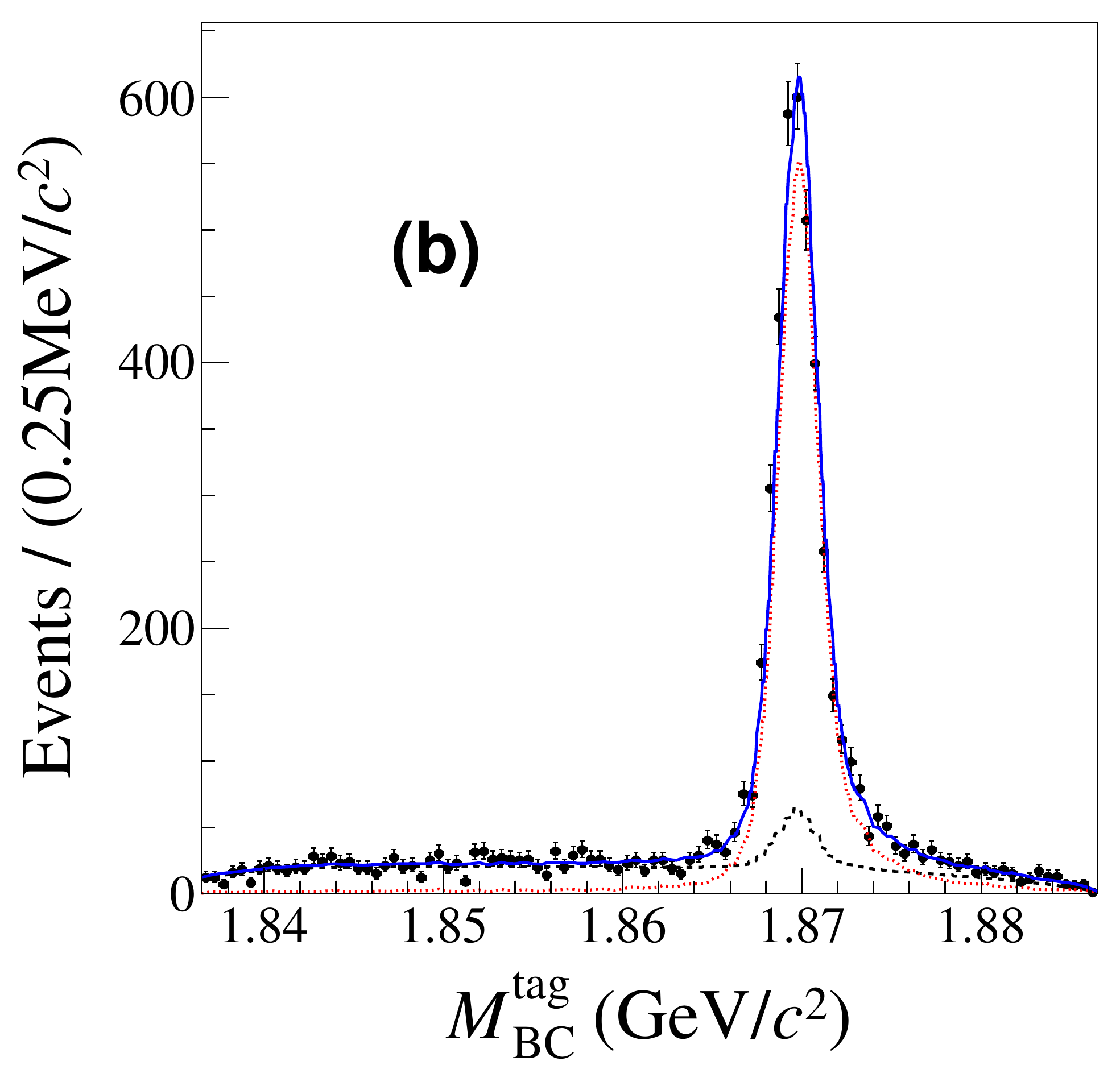}
    \includegraphics[width=6.cm]{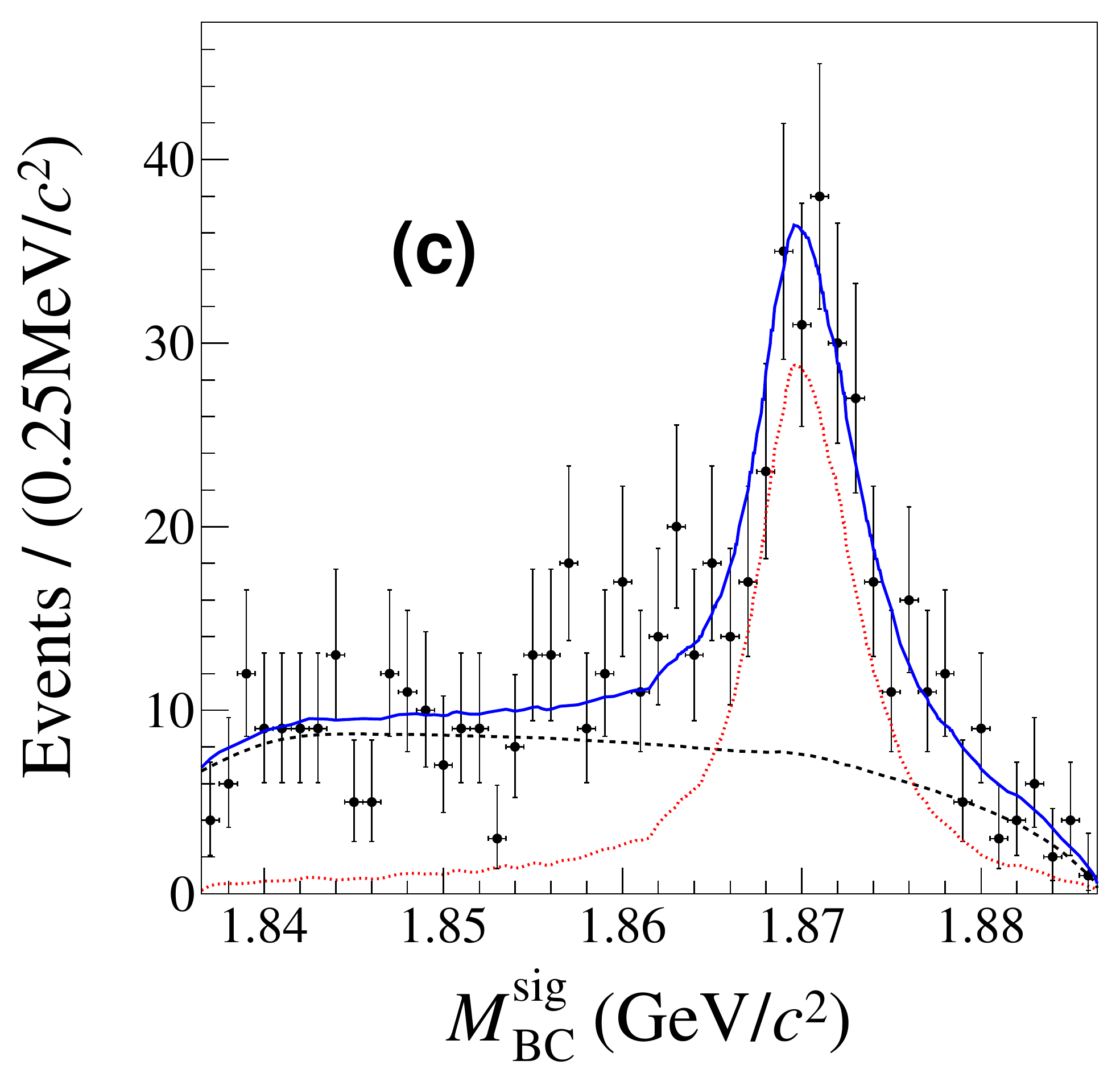}
    \includegraphics[width=6.cm]{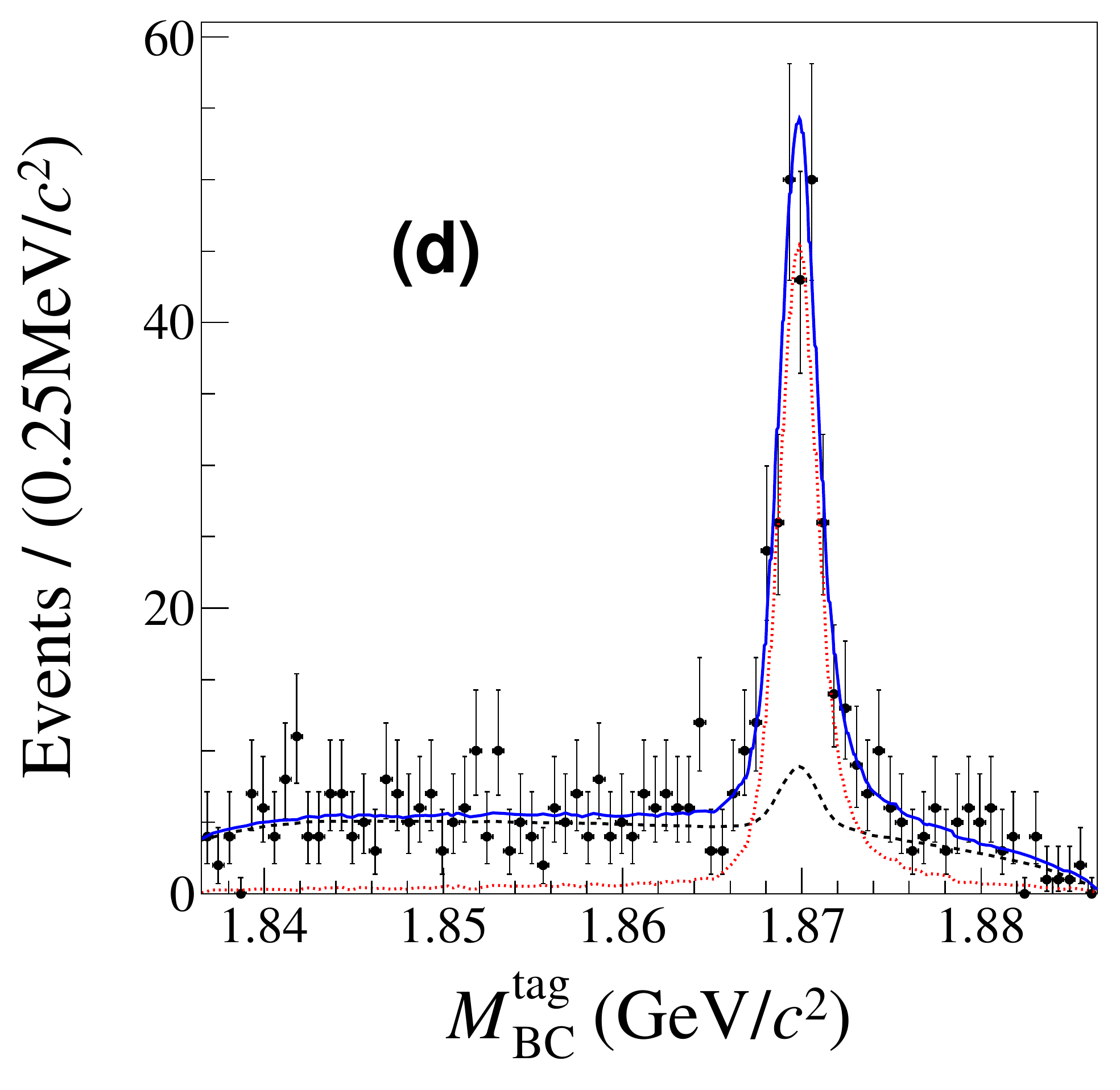}
\caption{The projections of the 2D fit of $M^{\rm sig}_{\rm BC}$ versus $M^{\rm tag}_{\rm BC}$ for data in the $K_S^0$ signal region (a)(b) and sideband region (c)(d). The black points with error bars are data. The blue solid lines are the total fits. The red dotted and black dashed lines are the fitted signal and background, respectively. }
\label{fig:DT_yield}
\end{center}
\end{figure*}

Once a tag mode is identified, we search for the signal decay $D^+ \to K_S^0\pi^+\pi^0\pi^0$ on the recoiling side using the condition $\Delta{E_{\rm sig}}\!\in\!\left[-0.063, 0.034\right]$~GeV. An unbinned 2D maximum likelihood fit is used to get the DT yield. In addition to the signal and background PDFs in appendix~\ref{app:2dfit}, one additional PDF based on a MC simulated shape is employed to describe the peaking background from $D^+\to K_S^0K_S^0\pi^+$. The corresponding yield is fixed to the estimation from the MC simulation. In order to estimate the combinatorial $\pi^+\pi^-$ backgrounds from $K_S^0$ reconstruction, we define the $K_S^0$ sideband region by $20\!<\!|M_{\pi^+\pi^-}\!-\!M_{K_s^0}|\!<\!44$~MeV$/c^2$ and perform the 2D fit in the $K_S^0$ signal and the sideband region, respectively. By subtracting the sideband contribution, the DT yield is calculated by
\begin{equation}
    N_{\rm total}^{\rm DT}=N^{\rm DT}_{K^0_S,{\rm sig}}-\cfrac{1}{2}N^{\rm DT}_{K^0_S,{\rm side}},
     \label{eq-DT}
\end{equation}
where $N^{\rm DT}_{K^0_S,{\rm sig}}$ and $N^{\rm DT}_{K^0_S,{\rm side}}$ denote the fitted yields in the $K_S^0$ signal and sideband regions, which are $3812\pm74$ and $266\pm23$, respectively. This relation has been verified by a large MC sample. Finally, the DT yield is obtained to be $3679\pm75$ and the fit results are shown in figure~\ref{fig:DT_yield}. Using a similar method for the signal MC samples, the DT efficiencies for various tag modes are determined and listed in table~\ref{tab:ST_yield}.

After correcting for the differences in $\pi^\pm$ tracking, PID and $\pi^0$ reconstruction efficiencies between data and MC simulation, we determine the BF to be $\mathcal{B}(D^+ \to K_S^0\pi^+\pi^0\pi^0)=(2.888\pm0.058_{\rm stat.}\pm0.069_{\rm syst.})\%$.

Most systematic uncertainties related to the efficiency of reconstructing the $D^\pm$ mesons on the tag side cancel due to the DT method. The following sources are taken into consideration to evaluate the systematic uncertainties in the BF measurement.
\begin{itemize}
    \item ST yield. The uncertainty of the total yield of the ST $D^\pm$ mesons has previously been estimated to be 0.5$\%$~\cite{BESIII:2016gbw,BESIII:2016hko,BESIII:2018nzb}, and is mainly due to the fits to the $M_{\rm BC}$ distributions of ST candidates.
    \item Tracking and PID efficiencies. The data-MC efficiency ratios for $\pi^+$ tracking and PID efficiencies are determined to be 1.001$\pm$0.001 and 0.998$\pm$0.001 for this decay channel by studying DT $D\bar{D}$ hadronic events. After correcting the MC efficiencies to data by these factors, the statistical uncertainties of the correction parameters are assigned to be the systematic uncertainties, which are 0.1\% for both $\pi^+$ tracking and PID.
    \item $K^0_S$ reconstruction. This systematic uncertainty is estimated from the measurements of $J/\psi\to K^*(892)^\mp K^\pm$ and $J/\psi\to \phi K_S^0K^\pm\pi^\mp$ control samples \cite{BESIII:2015jmz} and found to be 1.6\% per $K_S^0$.
    \item $\pi^0$ reconstruction. The data-MC efficiency ratio for $\pi^0$ reconstruction is determined to be 0.994$\pm$0.007 by using the $D\bar{D}$ hadronic decay samples of $D^0\to K^-\pi^+$, $K^-\pi^+\pi^+\pi^-$ versus $\bar{D}^0\to K^+\pi^-\pi^0$, $K_S^0\pi^0$. After correcting the efficiency by this factor for each $\pi^0$, we assign 0.7\% as the systematic uncertainty arising from the reconstruction of each $\pi^0$.
    \item MC sample size. The uncertainty of the limited MC sample size is given by $\sqrt{\sum_\alpha{(\frac{f_\alpha\delta_{\epsilon_\alpha}}{\epsilon_\alpha})^2}}$, where $f_\alpha$ is the tag yield fraction, $\epsilon_\alpha$ is the average DT efficiency of tag mode $\alpha$ and $\delta_{\epsilon_\alpha}$ is the uncertainty of $\epsilon_\alpha$. The corresponding uncertainty is determined to be 0.6\%.
    \item Quoted BFs. In this measurement, the BFs of the daughter particles are quoted from the PDG~\cite{PDG}, which are $\mathcal{B}(\pi^0\to \gamma\gamma)=(98.82\pm 0.04)\%$ and $\mathcal{B}(K^0_S\to \pi^+\pi^-)=(69.20\pm 0.05)\%$. The associated uncertainty is assigned to be 0.1\%.
    \item Amplitude model. The uncertainty from the amplitude model is determined by varying the amplitude model parameters based on their error matrix 600 times. A Gaussian function is used to fit the distribution of 600 DT efficiencies and the fitted width divided by the mean value is taken as the systematic uncertainty, which is 0.6\%.
    \item 2D fit. The signal and background shapes as well as the estimation of the size of the peaking background are the possible sources of uncertainty from the 2D fit. We vary the mean and width of the smeared Gaussian by $\pm1\sigma$ for the signal shape and the ARGUS end-point by $\pm$0.2~MeV/$c^2$ for the  background shape. Considering the uneven distribution for combinatorial $\pi^+\pi^-$ backgrounds of $K_S^0$ reconstruction, we also vary the $\frac{1}{2}$ factor in Eq.~(\ref{eq-DT}) according to its uncertainty from MC simulation. For the peaking background $D^+\to K_S^0K_S^0\pi^+$, whose yield is fixed in the 2D fit, we vary the quoted BF of this decay by $\pm1 \sigma$. The quadratic sum of the relative BF changes, 0.5\%, is assigned to be the systematic uncertainty for the 2D fit.
    \item $\Delta E_{\rm sig}$ requirement. Considering the possible difference between data and MC simulation, we examine the $\Delta E_{\rm sig}$ cut efficiency after smearing a double-Gaussian function for signal MC sample and we take the change of this efficiency to be the systematic uncertainty, which is 0.4\%.
\end{itemize}

All the systematic uncertainties are summarized in table~\ref{BF-Sys}. Adding them in quadrature results in a total systematic uncertainty of 2.6\% in the BF measurement.

\begin{table}[htbp]
  \centering
  \begin{tabular}{|l c|}
  \hline
    Source            & Uncertainty (\%) \\
    \hline
    ST yield                &0.5\\
    Tracking efficiency     &0.1\\
    PID efficiency          &0.1\\
    $K_S^0$ reconstruction  &1.6\\
    $\pi^0$ reconstruction  &1.4\\
    MC sample size          &0.6\\
    Quoted BFs    &0.1\\
    Amplitude model         &0.6\\
    2D fit                  &0.5\\
    $\Delta E_{\rm sig}$ requirement  &0.4\\
    \hline
    Total                   &2.4\\
    \hline
  \end{tabular}
    \caption{Systematic uncertainties in the BF measurement.}
  \label{BF-Sys}
\end{table}

\section{Summary}
Using an $e^+e^-$ collision data sample with an integrated luminosity of 2.93~$\rm{fb}^{-1}$ collected by the BESIII detector at $\sqrt{s}=3.773$~GeV, an amplitude analysis of $D^+\to K_S^0\pi^+\pi^0\pi^0$ is performed for the first time. The results for phases and FFs of different intermediate processes are listed in table~\ref{tab:signi}. With the detection efficiency obtained from a signal MC sample, which is generated based on our amplitude analysis model, the BF is determined to be $\mathcal{B}(D^+\to K_S^0\pi^+\pi^0\pi^0) = (2.888\pm0.058_{\rm stat.}\pm0.069_{\rm syst.})\%$. It is consistent with the previous BESIII result $(2.904\pm0.062_{\rm stat.}\pm0.087_{\rm syst.})\%$~\cite{BESIII:2022mji} within $1\sigma$, where the detection efficiency was simulated using mixed-signal MC samples.

\begin{table}[htbp]
  \centering   
  \begin{tabular}{|l r@{$\pm$}c@{$\pm$}c |}
    \hline
    Intermediate process  &\multicolumn{3}{c}{BF $(\times 10^{-3})$}  \vline\\
    \hline
    $D^+\to K_S^0a_1(1260)^+[S](\to \rho^+\pi^0)$
    &8.66  &1.04  &1.24 \\ \hline
    $D^+\to K_S^0a_1(1260)^+(\to f_0(500)\pi^+)$
    &1.00  &0.33  &0.55 \\
    \hline
    $D^+\to \bar{K}_1(1400)^0[S](\to \bar{K}^{*0}\pi^0)\pi^+$
    &1.73  &0.34  &0.09 \\
    $D^+\to \bar{K}_1(1400)^0[D](\to \bar{K}^{*0}\pi^0)\pi^+$
    &0.68  &0.16  &0.07 \\
    $D^+\to \bar{K}_1(1400)^0(\to \bar{K}^{*0}\pi^0)\pi^+$
    &2.32  &0.36  &0.13 \\
    \hline
    $D^+[S]\to \bar{K}^{*0}\rho^+$
    &9.20  &0.80  &0.45 \\
    $D^+[P]\to \bar{K}^{*0}\rho^+$
    &0.49  &0.17  &0.03 \\
    $D^+\to \bar{K}^{*0}\rho^+$
    &9.70  &0.81  &0.47 \\ \hline
    $D^+[S]\to \bar{K}^{*0}(\pi^+\pi^0)_V$
    &2.63  &0.57  &0.30 \\ \hline
    $D^+\to K_S^0(\rho^+\pi^0)_P$
    &4.75  &0.46  &0.14 \\
    \hline
  \end{tabular}
  \caption{The BFs for intermediate processes with the final state $D^+ \to K_S^0\pi^+\pi^0\pi^0$. The uncertainties are statistical and systematical respectively. Here, $\bar{K}^{*0}$ decays to $K_S^0\pi^0$, $\rho^+$ decays to $\pi^+\pi^0$, and $f_0(500)$ decays to $\pi^0\pi^0$.}
  \label{BF-FF}
\end{table}

We find $D^+\to \bar{K}^{*0}\rho^+$ and $D^+\to K_S^0a_1(1260)^+[S](\to \rho^+\pi^0)$ dominate in $D^+\to K_S^0\pi^+\pi^0\pi^0$ with FFs of $(33.6\pm2.7_{\rm stat.}\pm1.4_{\rm syst.})\%$ and $(30.0\pm3.6_{\rm stat.}\pm4.2_{\rm syst.})\%$, respectively, and obtain the BFs for intermediate processes presented in table~\ref{BF-FF}.

The absolute BF for $D^+\to \bar{K}^{*0}\rho^+$ is determined to be $(5.82\pm0.49_{\rm stat.}\pm0.28_{\rm syst.})\%$, which is consistent with the MARK III result $(4.8\pm1.2_{\rm stat.}\pm1.4_{\rm syst.})\%$~\cite{MARK-III:1991fvi} within $1\sigma$ but much more precise. The measured BF of $D^+\to K_S^0a_1(1260)^+[S](\to \rho^+\pi^0)$ is also consistent with the previous BESIII result~\cite{BESIII:2019ymv} within 1.5$\sigma$. We also observe an obvious $D^+\to \bar{K}_1(1400)^0(\to \bar{K}^{*0}\pi^0)\pi^+$ signal, but no $D^+\to K_1(1270)^+\pi^0$, where the significance is 3.9$\sigma$. This phenomenon is consistent with the theoretical prediction~\cite{Cheng:2003bn} and similar to that in $D^+\to K_S^0\pi^+\pi^+\pi^-$, where the FF of $D^+\to K_1(1400)\pi$ is about 10 times that of $D^+\to K_1(1270)\pi$~\cite{BESIII:2019ymv}. The specific BFs of $D\to K_1(1270)\pi$ and $D\to K_1(1400)\pi$ from amplitude analyses can provide inputs to further investigations of the mixing between these two axial-vector kaon mesons~\cite{Cheng:2011pb}.  

\acknowledgments
The BESIII Collaboration thanks the staff of BEPCII and the IHEP computing center for their strong support. This work is supported in part by National Key R\&D Program of China under Contracts Nos. 2020YFA0406400, 2020YFA0406300; National Natural Science Foundation of China (NSFC) under Contracts Nos. 11635010, 11735014, 11835012, 11935015, 11935016, 11935018, 11961141012, 12022510, 12025502, 12035009, 12035013, 12061131003, 12192260, 12192261, 12192262, 12192263, 12192264, 12192265, 12221005, 12225509, 12235017; the Chinese Academy of Sciences (CAS) Large-Scale Scientific Facility Program; the CAS Center for Excellence in Particle Physics (CCEPP); CAS Key Research Program of Frontier Sciences under Contracts Nos. QYZDJ-SSW-SLH003, QYZDJ-SSW-SLH040; 100 Talents Program of CAS; The Institute of Nuclear and Particle Physics (INPAC) and Shanghai Key Laboratory for Particle Physics and Cosmology; ERC under Contract No. 758462; European Union's Horizon 2020 research and innovation programme under Marie Sklodowska-Curie grant agreement under Contract No. 894790; German Research Foundation DFG under Contracts Nos. 443159800, 455635585, Collaborative Research Center CRC 1044, FOR5327, GRK 2149; Istituto Nazionale di Fisica Nucleare, Italy; Ministry of Development of Turkey under Contract No. DPT2006K-120470; National Research Foundation of Korea under Contract No. NRF-2022R1A2C1092335; National Science and Technology fund of Mongolia; National Science Research and Innovation Fund (NSRF) via the Program Management Unit for Human Resources \& Institutional Development, Research and Innovation of Thailand under Contract No. B16F640076; Polish National Science Centre under Contract No. 2019/35/O/ST2/02907; The Swedish Research Council; U. S. Department of Energy under Contract No. DE-FG02-05ER41374
\clearpage
\appendix
\section{Two-dimensional fit on $M^{\rm sig}_{\rm BC}$ versus $M^{\rm tag}_{\rm BC}$}
\label{app:2dfit}
\begin{figure*}[htp]
\begin{center}
    \includegraphics[width=6.8cm]{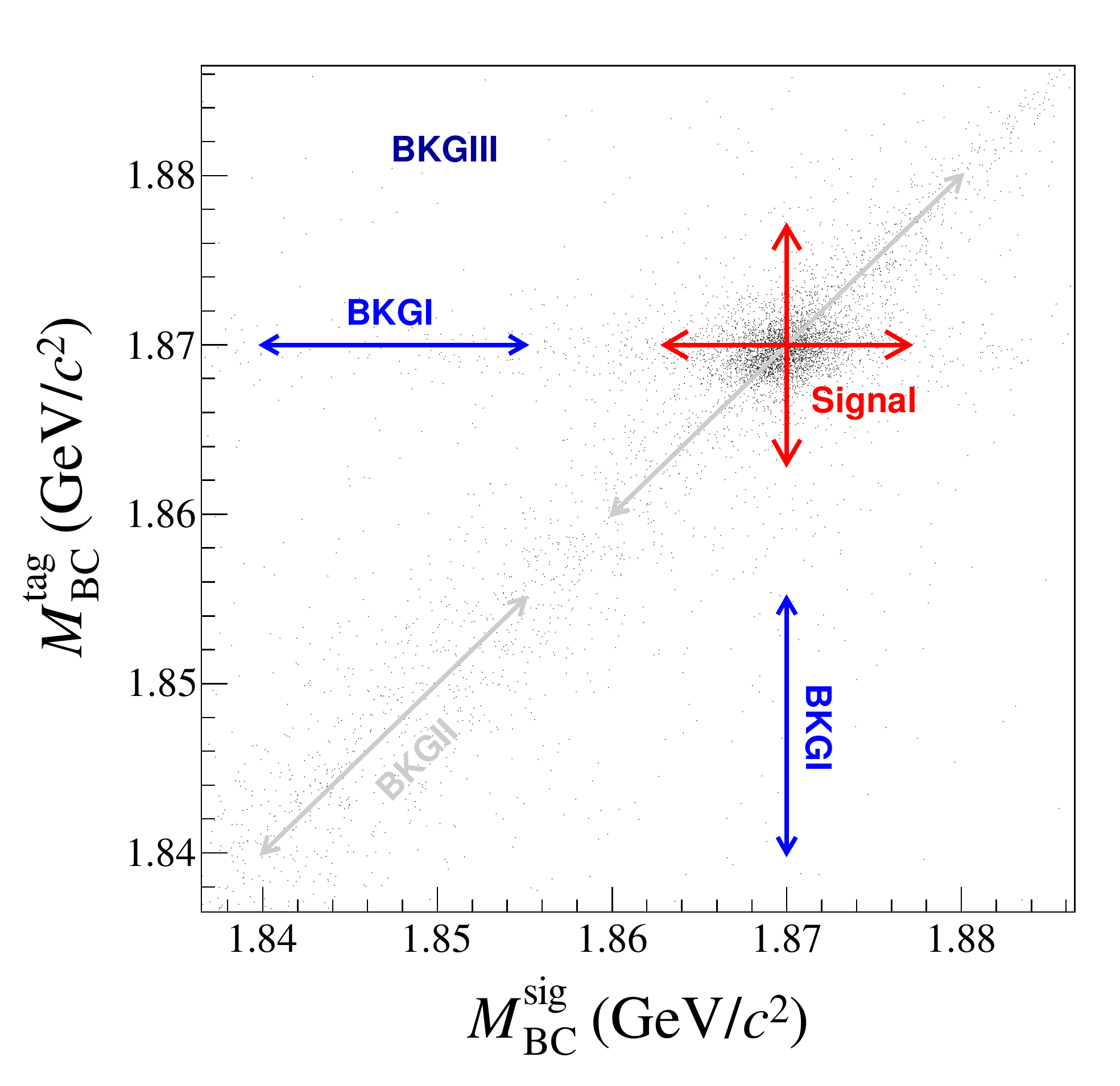}
\caption{The 2D $M^{\rm tag}_{\rm BC}$ versus $M^{\rm sig}_{\rm BC}$ distributions for data and different PDFs.}
\label{scatter}
\end{center}
\end{figure*}

The signal yield of DT candidates is determined by fitting to the 2D $M^{\rm tag}_{\rm BC}$ versus $M^{\rm sig}_{\rm BC}$ distribution. Signal events with both the tag side and signal side reconstructed correctly should concentrate around $M_{\rm BC}^{\rm sig} = M_{\rm BC}^{\rm tag} = M_{D^+}$, where $M_{D^+}$ is the known ${D^+}$ mass. Besides signal events, we define three kinds of background. Candidates with correctly reconstructed $D^+$(or $D^-$) and incorrectly reconstructed $D^-$(or $D^+$) are BKGI, which appear around the lines $M_{\rm BC}^{\rm sig}$ or $M_{\rm BC}^{\rm tag} = M_{D^+}$. Other candidates appearing around the diagonal are mainly from the $D^0\bar{D^0}$ wrong-combination and the $e^+e^-\to q\bar{q}$ processes (BKGII). The rest of the flat backgrounds mainly comes from candidates reconstructed incorrectly on both sides (BKGIII). Figure~\ref{scatter} shows the distributions of these PDFs. Here we list the probability density functions for different components in the fit:
\begin{itemize}
    \item Signal: $s(x,y)$,
    \item BKGI: $b_1(x)\cdot {\rm ARGUS}(y;m_0,c,p)+b_2(y)\cdot {\rm ARGUS}(x;m_0,c,p)$,
    \item BKGII: ${\rm ARGUS}((x+y)/\sqrt{2};m_0,c,p)\cdot g((x-y)/\sqrt{2})$,
    \item BKGIII: ${\rm ARGUS}(x;m_0,c,p)\cdot {\rm ARGUS}(y;m_0,c,p)$.
\end{itemize}

The signal shape $s(x,y)$ is described by the 2D MC-simulated shape convolved with a 2D Gaussian. For BKGI, $b_{1,2}(x,y)$ is described by the one-dimensional~(1D) MC-simulated shape convoluted with a Gaussian, ${\rm ARGUS}(x,y)$ is the ARGUS function~\cite{ARGUS:1990hfq}. The parameters of the convoluted Gaussian functions are obtained by a 1D fit to $M_{\rm BC}$ on the signal and tag side respectively, and are fixed in the 2D fit. For BKGII, it is an ARGUS function in the diagonal axis multiplied by a Gaussian in the anti-diagonal axis. For BKGIII, it is an ARGUS function in $M_{\rm BC}^{\rm sig}$ multiplied by an ARGUS function in $M_{\rm BC}^{\rm tag}$. In the fit, the parameters $m_0$ and $p$ for the ARGUS function are fixed at 1.8865 GeV and 0.5, respectively.

\clearpage
\section{Other tested intermediate processes}
\label{app:others}
Some other tested amplitudes with significance less than 5$\sigma$ are listed below. The significance is given in brackets. The resonance $f_0(980)$ only decays to $\pi^0\pi^0$.

\begin{itemize}
    \item Cascade amplitudes
    \begin{itemize}
        \item[-] $D^+\to K_S^0a_1(1260)^+[D](\to \rho^+\pi^0$) ($<1\sigma$)        
        \item[-] $D^+\to K_S^0a_1(1260)^+(\to f_0(980)\pi^+$) ($3.5\sigma$)  
        \item[-] $D^+\to K_1(1270)^+[S](\to K_S^0\rho^+)\pi^0$ ($1.7\sigma$)
        \item[-] $D^+\to K_1(1270)^+[D](\to K_S^0\rho^+)\pi^0$ ($3.2\sigma$)
        \item[-] $D^+\to K_1(1270)^+(\to K_S^0\rho^+)\pi^0$ ($3.9\sigma$)
        \item[-] $D^+\to K(1460)^0(\to \bar{K}^{*0}\pi^0)\pi^+$ ($3.3\sigma$)  
        \item[-] $D^+[D]\to \bar{K}^{*0}\rho^+$ ($<1\sigma$)

   \end{itemize}
   \item Three-body amplitudes
   \begin{itemize}
       \item[-] $D^+\to (K_S^0\pi^0)_{S-wave}f_0(980)$ ($<1\sigma$) 
       \item[-] $D^+\to (\bar{K}^{*0}\pi^+)_P\pi^0$ ($3.1\sigma$) 
       \item[-] $D^+\to (\bar{K}^{*0}\pi^0)_P\pi^+$ ($<1\sigma$)   
       \item[-] $D^+\to (K_S^0\rho^+)_P\pi^0$ ($<1\sigma$)
       \item[-] $D^+\to K_S^0(\rho^+\pi^0)_A[S]$ ($3.0\sigma$)
       \item[-] $D^+\to K_S^0(\rho^+\pi^0)_A[D]$ ($3.6\sigma$)      
       \item[-] $D^+\to (\bar{K}^{*0}\pi^0)_A[S]\pi^+$ ($3.0\sigma$) 
       \item[-] $D^+\to (\bar{K}^{*0}\pi^0)_A[D]\pi^+$ ($<1\sigma$)  
       \item[-] $D^+\to (K_S^0\rho^+)_A[D]\pi^0$ ($2.9\sigma$)    
       \item[-] $D^+\to (K_S^0\pi^0)_{S-wave}\rho^+$ ($4.2\sigma$) 
       \item[-] $D^+\to \bar{K}^{*0}(\pi^+\pi^0)_S$ ($3.2\sigma$) 
       \item[-]  $D^+\to (\bar{K}^{*0}\pi^0)_V\pi^+$ ($2.2\sigma$)   
       \item[-] $D^+\to K_S^0(\pi^0\rho^+)_V$ ($<1\sigma$)           
       \item[-] $D^+[S]\to (K_S^0\pi^0)_V\rho^+$ ($3.5\sigma$)       
       \item[-] $D^+[P]\to (K_S^0\pi^0)_V\rho^+$ ($<1\sigma$)        
       \item[-] $D^+[D]\to (K_S^0\pi^0)_V\rho^+$ ($3.4\sigma$)       
       \item[-] $D^+[P]\to \bar{K}^{*0}(\pi^+\pi^0)_V$ ($2.2\sigma$) 
       \item[-] $D^+[D]\to \bar{K}^{*0}(\pi^+\pi^0)_V$ ($<1\sigma$)  
   \end{itemize}
   \item Four-body non-resonance amplitudes
   \begin{itemize}
       \item[-] $D^+\to (K_S^0\pi^0)_{S-wave}(\pi^+\pi^0)_S$ ($<1\sigma$)        
       \item[-] $D^+\to (K_S^0\pi^+)_{S-wave}(\pi^0\pi^0)_S$  ($<1\sigma$)    
       \item[-] $D^+\to K_S^0((\pi^+\pi^0)_S\pi^0)_A$ ($1.9\sigma$)  
       \item[-] $D^+\to K_S^0((\pi^0\pi^0)_S\pi^+)_A$ ($1.6\sigma$)  
       \item[-] $D^+\to ((K_S^0\pi^+)_{S-wave}\pi^0)_A\pi^0$ ($1.2\sigma$)        
       \item[-] $D^+\to ((K_S^0\pi^0)_{S-wave}\pi^+)_A\pi^0$ ($1.8\sigma$)    
       \item[-] $D^+\to ((K_S^0\pi^0)_{S-wave}\pi^0)_A\pi^+$ ($<1\sigma$)       
       \item[-] $D^+\to (K_S^0(\pi^+\pi^0)_S)_A\pi^0$ ($2.1\sigma$)  
       \item[-] $D^+\to (K_S^0(\pi^0\pi^0)_S)_A\pi^+$ ($<1\sigma$)   
       \item[-] $D^+\to (K_S^0\pi^0)_{S-wave}(\pi^+\pi^0)_V$ ($2.3\sigma$)  
       \item[-] $D^+\to (K_S^0\pi^0)_V(\pi^+\pi^0)_S$ ($2.2\sigma$)  
       \item[-] $D^+[S]\to (K_S^0\pi^0)_V(\pi^+\pi^0)_V$ ($<1\sigma$) 
       \item[-] $D^+[P]\to (K_S^0\pi^0)_V(\pi^+\pi^0)_V$ ($2.4\sigma$) 
       \item[-] $D^+[D]\to (K_S^0\pi^0)_V(\pi^+\pi^0)_V$ ($2.0\sigma$) 
       \item[-] $D^+\to (K_S^0\pi^+)_T(\pi^0\pi^0)_S$ ($2.4\sigma$)  
       \item[-] $D^+\to (K_S^0\pi^0)_T(\pi^+\pi^0)_S$ ($3.6\sigma$)  
       \item[-] $D^+\to (K_S^0\pi^+)_S(\pi^0\pi^0)_T$ ($2.1\sigma$)  
       \item[-] $D^+\to (K_S^0\pi^0)_S(\pi^+\pi^0)_T$ ($1.9\sigma$)  
   \end{itemize}
\end{itemize}

\section{The interference between processes}
The interference between processes, calculated by Eq.~(\ref{inter}).

\label{app:inter}
\begin{table}[htbp]\small
  \centering
  \begin{tabular}{|c| r r r r r r r|}
    \hline
    &II &III &IV &V &VI &VII &VIII\\
    \hline
    I    & 4.88 &-1.09 &-0.29 &15.43 &-0.00 &-4.39 &0.07 \\
    II   & &-1.37 & 0.05 & 5.95 & 0.00 &-2.41 & 0.01 \\
    III  & & &-0.22 &-9.81 &-0.00 & 7.28 &-0.27 \\
    IV   & & & & 1.01 & 0.00 &-0.17 &-0.19 \\
    V    & & & & &-0.00 &-21.45 &9.37 \\
    VI   & & & & & & 0.00 &-0.00 \\
    VII  & & & & & & &-2.98 \\
    \hline
  \end{tabular}
    \caption{Interference between processes, in unit of \%.\\
    I~~~~  $D^+\to K_S^0a_1(1260)^+, a_1(1260)^+[S]\to \rho^+\pi^0$,\\
    II~~~  $D^+\to K_S^0a_1(1260)^+, a_1(1260)^+\to f_0(500)\pi^+$,\\
    III~~  $D^+\to \bar{K}_1(1400)^0\pi^+, \bar{K}_1(1400)^0[S]\to \bar{K}^{*0}\pi^0$,\\
    IV~~   $D^+\to \bar{K}_1(1400)^0\pi^+, \bar{K}_1(1400)^0[D]\to \bar{K}^{*0}\pi^0$,\\
    V~~~   $D^+[S]\to \bar{K}^{*0}\rho^+$,\\
    VI~~   $D^+[P]\to \bar{K}^{*0}\rho^+$,\\
    VII~   $D^+[S]\to \bar{K}^{*0}(\pi^+\pi^0)_V$,\\
    VIII   $D^+\to K_S^0(\rho^+\pi^0)_P$.}
      \label{tab:inter}
\end{table}
\clearpage
\bibliographystyle{JHEP}
\bibliography{references}
\clearpage
\large
The BESIII Collaboration\\
\normalsize
\\M.~Ablikim$^{1}$, M.~N.~Achasov$^{5,b}$, P.~Adlarson$^{75}$, X.~C.~Ai$^{81}$, R.~Aliberti$^{36}$, A.~Amoroso$^{74A,74C}$, M.~R.~An$^{40}$, Q.~An$^{71,58}$, Y.~Bai$^{57}$, O.~Bakina$^{37}$
% ...
\\
\vspace{0.2cm}

\vspace{0.2cm} {\it
\noindent$^{1}$ Institute of High Energy Physics, Beijing 100049, People's Republic of China\\
% ...
$^{5}$ Budker Institute of Nuclear Physics SB RAS (BINP), Novosibirsk 630090, Russia\\
% ...
$^{36}$ Johannes Gutenberg University of Mainz, Johann-Joachim-Becher-Weg 45, D-55099 Mainz, Germany\\
$^{37}$ Joint Institute for Nuclear Research, 141980 Dubna, Moscow region, Russia\\
% ...
$^{40}$ Liaoning Normal University, Dalian 116029, People's Republic of China\\
% ...
$^{57}$ Southeast University, Nanjing 211100, People's Republic of China\\
$^{58}$ State Key Laboratory of Particle Detection and Electronics, Beijing 100049, Hefei 230026, People's Republic of China\\
% ...
$^{71}$ University of Science and Technology of China, Hefei 230026, People's Republic of China\\
% ...
$^{74}$ University of Turin and INFN, (A)University of Turin, I-10125, Turin, Italy; (B)University of Eastern Piedmont, I-15121, Alessandria, Italy; (C)INFN, I-10125, Turin, Italy\\
$^{75}$ Uppsala University, Box 516, SE-75120 Uppsala, Sweden\\
% ...
$^{81}$ Zhengzhou University, Zhengzhou 450001, People's Republic of China\\

\vspace{0.2cm}
\noindent$^{a}$ Also at the Moscow Institute of Physics and Technology, Moscow 141700, Russia\\
$^{b}$ Also at the Novosibirsk State University, Novosibirsk, 630090, Russia\\
% ...
}

\end{document}